\documentclass[onecolumn]{aastex631}

\shorttitle{Deep HI Mapping of Stephan's Quintet and Its Neighborhood}
\shortauthors{Cheng Cheng et al.}
\begin{document}

\title{Deep HI Mapping of Stephan's Quintet and Its Neighborhood}

\correspondingauthor{Cong Kevin Xu}
\email{xucong@nao.cas.cn}

\author{Cheng Cheng}
\affiliation{Chinese Academy of Sciences South America Center for Astronomy, National Astronomical Observatories, CAS, Beijing 100101, China}
\affiliation{National Astronomical Observatories, Chinese Academy of Sciences, 20A Datun Road, Chaoyang District, Beijing 100101, People’s Republic of China}
\affiliation{CAS Key Laboratory of Optical Astronomy, National Astronomical Observatories, Chinese Academy of Sciences, Beijing 100101, China}
\author{Cong Kevin Xu}
\affiliation{Chinese Academy of Sciences South America Center for Astronomy, National Astronomical Observatories, CAS, Beijing 100101, China}
\affiliation{National Astronomical Observatories, Chinese Academy of Sciences, 20A Datun Road, Chaoyang District, Beijing 100101, People’s Republic of China}%
\author{P. N. Appleton}
\affiliation{Caltech/IPAC, MC 6-313, 1200 E. California Blvd., Pasadena, CA 91125, USA}
\author{P.-A. Duc}
\affiliation{Universit\'e de Strasbourg, CNRS, Observatoire astronomique de Strasbourg, UMR 7550, F-67000 Strasbourg, France}
%
%
\author{N.-Y. Tang}
\affiliation{Department of Astronomy, University of Massachusetts, Amherst, MA 01003, USA}
\author{Y.S. Dai}
\affiliation{Chinese Academy of Sciences South America Center for Astronomy, National Astronomical Observatories, CAS, Beijing 100101, China}
\author{J.-S. Huang}
\affiliation{Chinese Academy of Sciences South America Center for Astronomy, National Astronomical Observatories, CAS, Beijing 100101, China}
\affiliation{Center for Astrophysics \textbar\ Harvard \& Smithsonian, 60 Garden St., Cambridge, MA 02138 USA}
\author{U. Lisenfeld}
\affiliation{Dept. F\'isica Teórica y del Cosmos, Campus de Fuentenueva, Edificio Mecenas, Universidad de Granada, E-18071 Granada, Spain}
\affiliation{Instituto Carlos I de Física T\'orica y Computacional, Facultad de Ciencias, E-18071 Granada, Spain}
\author{F. Renaud}
\affiliation{Department of Astronomy and Theoretical Physics, Lund Observatory, Box 43, SE-221 00 Lund, Sweden}
\author{Chuan He}
\affiliation{Chinese Academy of Sciences South America Center for Astronomy, National Astronomical Observatories, CAS, Beijing 100101, China}
\author{Hai-Cheng Feng}
\affiliation{Yunnan Observatories, Chinese Academy of Sciences, Kunming 650011, Yunnan, People's Republic of China}
\affiliation{University of Chinese Academy of Sciences, Beijing 100049, People's Republic of China} 
\affiliation{Key Laboratory for the Structure and Evolution of Celestial Objects, Chinese Academy of Sciences, Kunming 650011, Yunnan, People's Republic of China}

\begin{abstract}
We carried out deep mapping observations of the atomic hydrogen (HI) 21 cm line emission in a field centered on the famous galaxy group Stephan’s Quintet (SQ), using the Five-hundred-meter Aperture Spherical Telescope (FAST) equipped with the 19-Beam Receiver. 
The final data cube reaches an HI column density sensitivity of $5 \sigma =  2.1\times 10^{17}$ cm$^{-2}$ per 20 km s$^{-1}$ channel with an angular resolution of  $4'.0$.
The discovery of a large diffuse feature of the HI emission in the outskirt of the intragroup medium of SQ was reported in a previous paper \citep{Xu2022}. Here we present a new study of the total HI emission of SQ and the detection of several neighboring galaxies, exploiting the high sensitivity and the large sky coverage of the FAST observations. A total HI mass of $\rm M_{HI} = 3.48  \pm 0.35 \times 10^{10}\; M_\sun$ is found for SQ, which is significantly higher than previous measurements in the literature. This indicates that, contrary to earlier claims, SQ is not HI deficient. The excessive HI gas is mainly found in the velocity ranges of 6200 - 6400 km s$^{-1}$ and  6800 - 7000 km s$^{-1}$, which was undetected in previous observations that are less sensitive than ours. Our results suggest that the ``missing HI" in compact groups may be hidden in the low-density diffuse neutral gas instead of in the ionized gas.
\end{abstract}

\keywords{Galaxy groups (597); Hickson compact group (729); Stefan’s Quintet  (1575)}

\section{Introduction} \label{sec:intro}
Galaxies are formed in hierarchical structures in the Universe. Galaxy-galaxy interaction plays an important role in the formation and evolution of galaxies. Compact groups, characterized by aggregates of 4 – 8 galaxies with space densities as high as those in cluster cores  \citep{Hickson1982},  represent a special class of interacting systems completely different from merging pairs that trigger the most extreme starbursts (star formation rate, SFR $> 100$ M$_\odot$ yr$^{-1}$) in the local Universe  \citep{Sanders1996}. Galaxies in compact groups show very diversified SFR, but none with detected SFR $>$ 100 M$_\odot$ yr$^{-1}$ \citep{Lenkic2016}. 
Their SFR depends sensitively on the HI abundance 
which is generally deficient \citep{VerdesMontenegro2001, Johnson2007}. Discovered as an aggregation of nebulae in 1877 \citep{Stephan1877}, SQ is arguably the best-studied compact group. It has been observed in nearly all accessible wavebands from X-rays, UV, visible, to IR and radio \citep{Allen1972, vanderhulst1981, Shostak1984,  Yun1997, Moles1997, Xu1999, Gao2000, Sulentic2001, Gallagher2001, Williams2002, Xu2003,  Lisenfeld2004, Xu2005, Trinchieri2005, Appleton2006}, and high-resolution observations of the SQ's infrared (IR) emission were included in the James-Webb Space Telescope's first public release \citep{Pontoppidan2022}. It has been found that the HI gas is all in the intragroup medium outside member galaxies in SQ \citep{Shostak1984, Williams2002}, and its spiral members show non-enhanced SFR \citep{Xu2005}. Most strikingly, \citet{Allen1972} discovered in the intragroup medium of SQ  a very large shock front of $\sim 40$ kpc in size which, among all known intergalactic shock fronts,  is only second to that of the radio relics caused by cluster mergers \citep{Ensslin2002}.  This shock was triggered by a high velocity ($\sim$1000 km s$^{-1}$) intruder (NGC 7318b) currently colliding into a debris field in the intragroup medium, the latter being a product of previous interactions among other member galaxies of the group \citep{Allen1972, vanderhulst1981, Moles1997, Sulentic2001}. SQ has provided an excellent laboratory for studies of various processes associated with high-speed galaxy collisions, such as the heating and cooling of warm H$_2$ gas powered by large-scale shocks \citep{Appleton2017, Guillard2022}, SFR quenching by turbulence in post-shock molecular gas \citep{Guillard2012}, and the formation of Green Valley galaxies in galaxy groups \citep{Cluver2013, Alatalo2014, Lisenfeld2017}. These studies provide important constraints to galaxy formation/evolution models, in particular for the high z Universe when galaxies have very low metallicity and may collide with each other more frequently. 

A comprehensive understanding of the properties of SQ depends crucially on how accurately we know its interaction history, which is very complex due to the relatively large number of participants ($N \geq 5$). Theoretical simulations have yielded inconclusive and controversial results. The fiducial model in \citet{Renaud2010} involves a number of events in the last half Gyr, including collisions between spiral galaxies NGC 7319 and NGC 7320c creating the so-called outer tail, between NGC 7319 and the early-type galaxy NGC 7318a creating the inner tail, and between the high-speed intruder NGC 7318b and the intragroup medium producing shocks in the gas stripped by the previous interactions. However, the hydrodynamical simulations in \citet{Hwang2012} favor a different scenario in which both the outer tail and inner tail were triggered simultaneously by a close encounter between NGC 7320c and NGC 7319 about $9\times10^8$ years ago. It provides a better fit to the current positions of galaxies and tidal tails, though cannot explain the observed age difference between star formation regions and star clusters in the outer and inner tails \citep{Xu2005, Fedotov2011}. It predicts that most gas is still inside galaxies which contradicts with observations \citep{Shostak1984, Williams2002}. In both models, the role played by NGC 7317, an early-type galaxy, is neglected because of a lack of observed connection to any other SQ members. Recently a reddish, low surface brightness stellar halo surrounding SQ was discovered in deep optical imaging observations \citep{Duc2018}, which corresponds spatially very well to the diffuse X-ray halo \citep{Trinchieri2005}, in particular around NGC 7317. Its size and morphology are consistent with the formation of the galaxy group more than a billion years ago. 

In order to better constrain the early history of SQ and explore the diffuse HI gas in and around the group, we carried out deep mapping observations of the diffuse HI emission in a region of  $\sim 30' \times 30'$ centered on SQ using the Five-hundred-meter Aperture Spherical Telescope (FAST) equipped with the 19-Beam Receiver \citep{2019SCPMA..6259502J, Jiang2020, 2020Innov...100053Q}. The HI is the least bound component of galaxies, so will be the easiest (and hence first) to be stripped off and spread around during interactions. Thus, the distribution of the diffuse HI and its velocity field can put new and crucial constraints to the interaction history models. 
The discovery of a large diffuse feature of the HI emission in the outskirt of intragroup medium of SQ was reported in a previous paper \citep{Xu2022}. In this paper we present other results from the same observations, with the focus on the contribution of the diffuse gas to the total HI mass in the central region of the intragroup medium, and the detection of several neighboring galaxies.

The paper is organized as follows: after this introduction, Section 2 describes the observations and the data reduction procedure.  Sections 3 and 4 present our results on the HI properties of SQ and of the neighboring galaxies, respectively. Section 5 is devoted to a discussion. Finally, a summary is given in Section 6. A comoving distance of 85 Mpc for SQ is adopted from NED\footnote[1]{The NASA/IPAC Extragalactic Database (NED) is operated by the Jet Propulsion Laboratory, California Institute of Technology, under contract with the National Aeronautics and Space Administration.}. For neighboring galaxies in the field of view of the FAST observations, a Hubble constant of $\rm H_0 = 70\; km\; s^{-1}\; Mpc^{-1}$ is assumed. 

\begin{figure*}
    \centering
    \includegraphics[width = 0.99\textwidth]{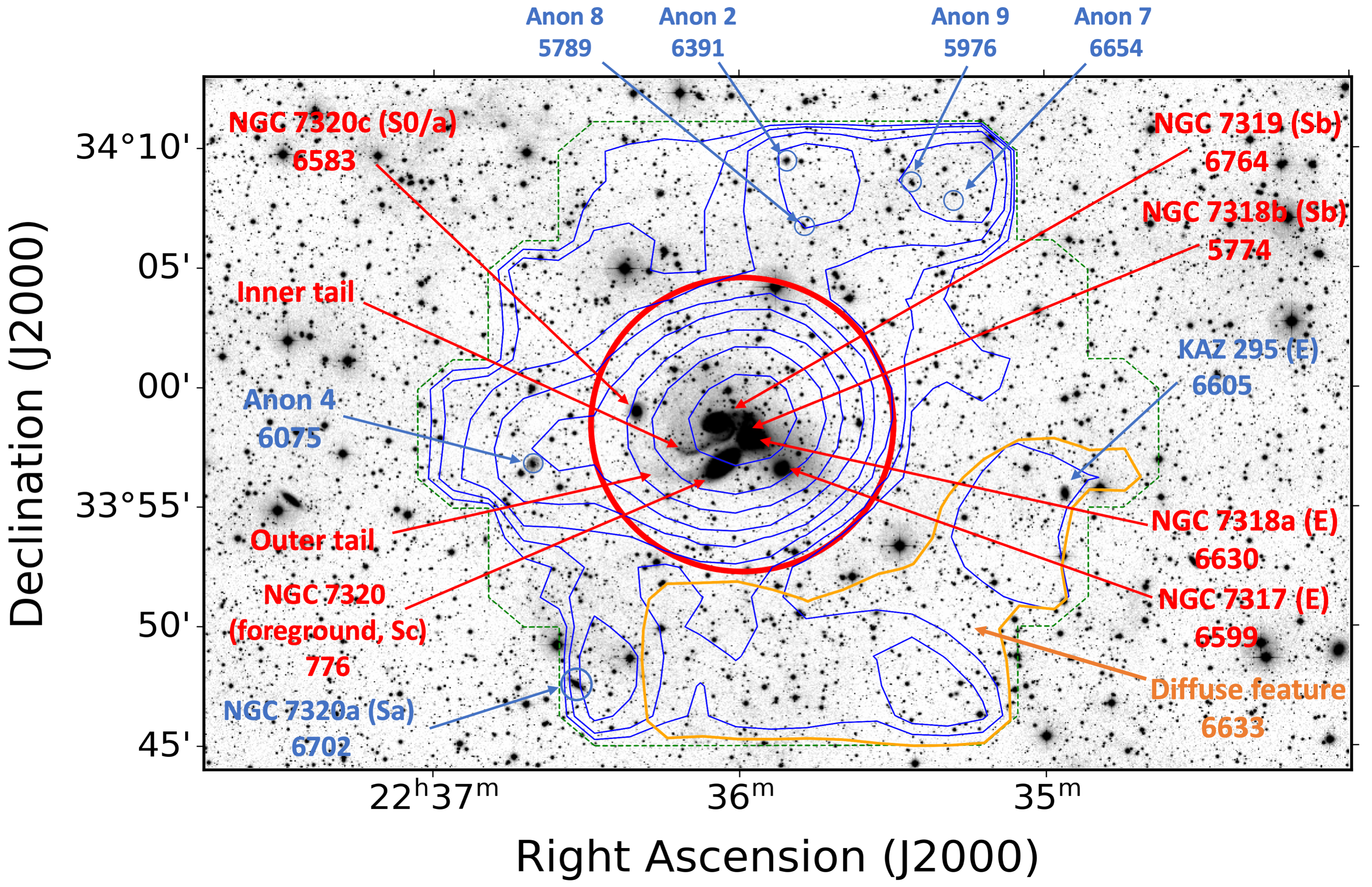}
    \caption{
    Contour map (in blue contours) of the integrated HI emission between 5600 - 7000 km s$^{-1}$  over-plotted on the r band image. 
    The contour levels are [4, 8, 16, 32, 64, ...]$\times 3.5\times 10^{17} \rm cm^{-2}$. The green boundary delineates the coverage of the FAST observations. The central red circle ($\rm D=12'$) shows the aperture with which the total HI spectrum presented in Figure \ref{totalHIspec} is extracted. The orange line delineates the diffuse HI feature whose system velocity is written in brown labels \citep{Xu2022}. Provided in the figure is also information for the member galaxies and main tidal features in SQ (in red labels), for neighboring galaxies detected by FAST (in blue labels) whose positions are marked by blue circles, and for neighboring galaxy KAZ~295 (in blue labels). For each source, the name, Hubble type (when available), and velocity in km s$^{-1}$ are listed.}
    \label{totalHImap}
\end{figure*}

\begin{figure*}
    \centering
    \includegraphics[width = 0.94\textwidth]{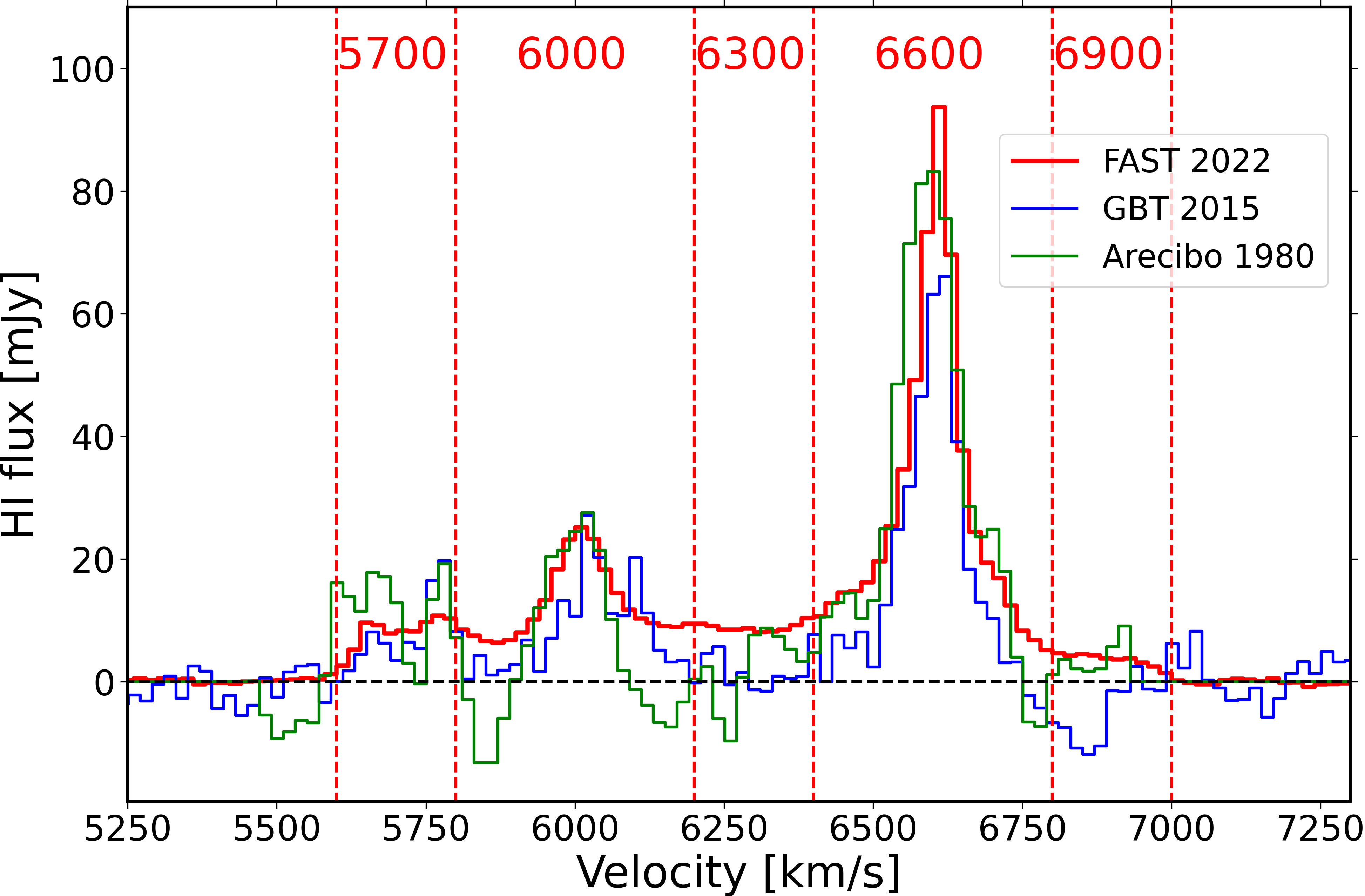}
    \caption{HI spectra of SQ obtained by FAST (red), Arecibo \citep[green,][]{1980ApJ...241L...1P} and GBT \citep[blue,][]{Borthakur2015}. The red vertical dashed lines mark the boundaries of the five kinematical components.
    We have smoothed the Arecibo and GBT spectra to a resolution of 20 km s$^{-1}$ for comparison, from their original resolutions of 8.8 km s$^{-1}$ and 10 km s$^{-1}$, respectively.
    }
    \label{totalHIspec}
\end{figure*}

\section{Observations and Data Reduction}
The FAST observations and the data reduction, including the sidelobe correction, are presented in detail in \citet{Xu2022}. Here we give only a brief recount:

{\bf Observations:} The observations were carried out using the FAST 19-beam receiver in the standard ON-OFF mode with a total observation time of 22.4 hours (including overheads). They consist of 16 pointings in a rectangular $4 \times 4$ grid and cover uniformly a region of $\sim 30'$ in size with 304 sky pixels (beam positions in the sky). The angular resolution is $2.9'$  and the separation between the nearest pixels is $1.4'$ in the right ascension (RA) direction and $1.2'$ in the declination (Decl.) direction, therefore the mapping satisfies the Nyquist sampling criterion. The observations have a central frequency of 1391.64 MHz and a frequency coverage of 1050 – 1450 MHz with a frequency resolution of 7.63 kHz ($\Delta v = 1.65$ \rm km s$^{-1}$).

{\bf Data reduction:} The spectral data of individual beams were reduced following a similar procedure as presented in \citet{Cheng2020}.  This includes: averaging of individual samplings, calibration (converting the digital counts to flux density in mJy beam$^{-1}$), ON-minus-OFF, standing-waves and baseline removal, and re-binning. In addition, the sidelobes are corrected utilizing the images of individual beams  \citep{Jiang2020} which provide information on the point spread functions (PSF’s) of the beams. The end product is a data cube containing 304 spectra covering the velocity range of  4600 – 7600 km s$^{-1}$ in $\Delta v = 20$ km s$^{-1}$ channels with a 1$\sigma$ noise of 0.16 mJy beam$^{-1}$. The corresponding HI column density sensitivity is $1\sigma =  1.2\times 10^{17}$ cm$^{-2}$ per channel. The velocity system is in the optical redshift convention and the local standard of rest (LSR) reference frame. The calibration uncertainty of FAST data is 10\% \citep{Jiang2020}.

{\bf Smoothing:} The data cube obtained above is highly redundant in the sense that a sky area of the size of a  single beam (D=$2.9'$) is covered by multiple beams (beam-separation: $1.4' \times 1.2'$). In order to take the advantage of the redundancy and improve the sensitivity for the diffuse emission, a Gaussian kernel of Full-Width-at-Half-Maximum (FWHM) $= 2'.8$ is applied to the data cube. This improves the HI column density sensitivity to  $1\sigma =  4.2\times 10^{16}$ cm$^{-2}$ per channel with a slightly degraded angular resolution of  $4'.0$.

The channel maps of the FAST observations are presented in Appendix A. 

\begin{figure*}
    \centering
    \includegraphics[width = 0.94\textwidth]{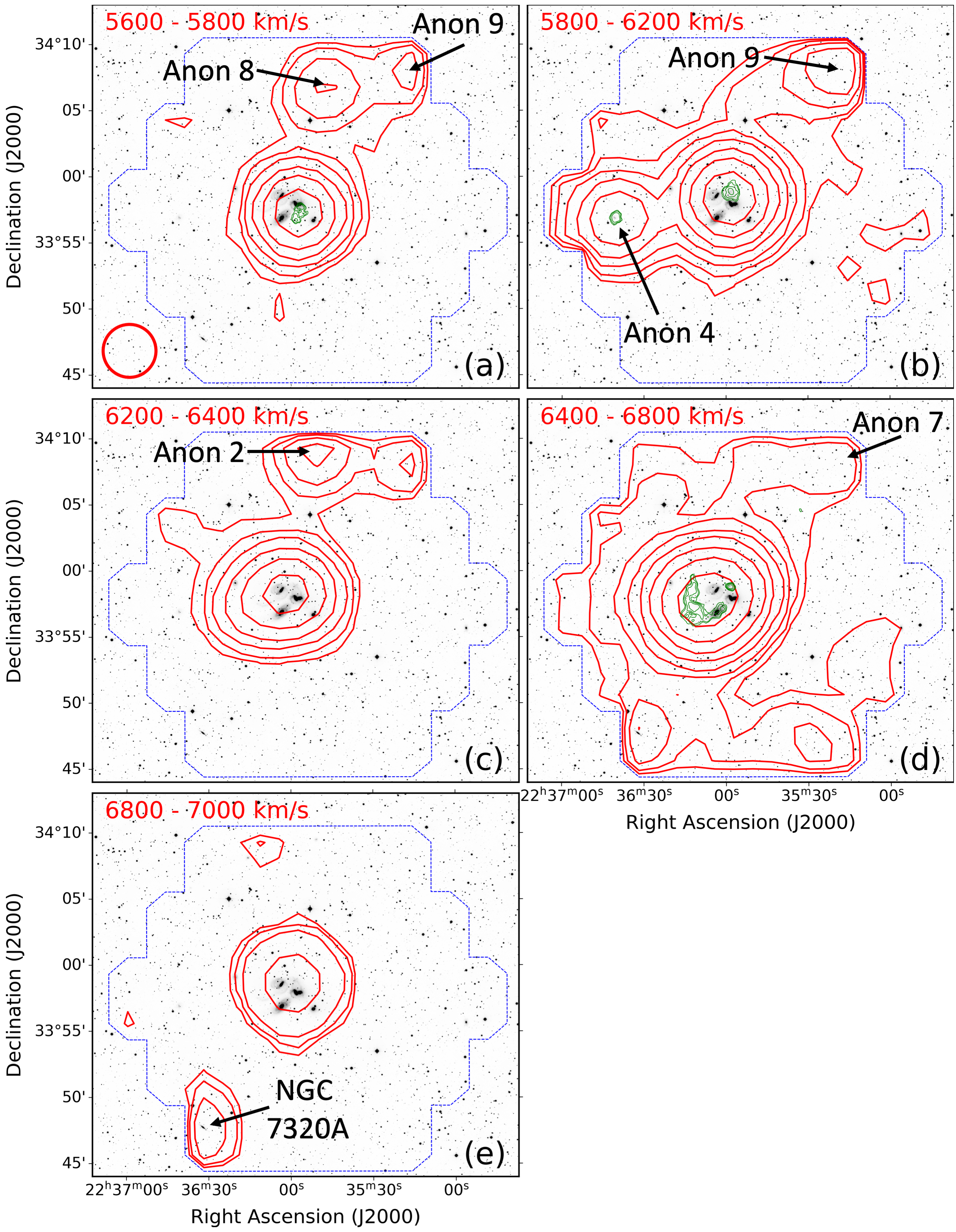}
    \caption{Contour maps of the integrated flux of the five kinematical components (see Figure \ref{totalHIspec}) over-plotted on the r-band image. The red contour levels are [4, 8, 16, 32, 64, ...] $\times \sigma$, where $ \sigma = \sqrt{\Delta v /20 \rm ~km~s^{-1}} \times 4.2\times 10^{16} \rm cm^{-2}$, and the velocity range of each component is provided in the corresponding panel. Marked on the maps are also the neighboring galaxies detected by the FAST observations (see Section 4). The red circle at the bottom-left corner of panel (a) shows the angular resolution of the contour maps (FWHM= $4'.0$).
    The green contours in panels (a), (b), and (d) represent the VLA moment 0 maps in velocity ranges of 5597–5789 km s$^{-1}$, 5959–6068 km s$^{-1}$, and 6475–6755 km s$^{-1}$, respectively. The contours levels are 5.8, 12, 18, 23, 29, 44, 58, and 73 $\times 10^{19} ~ \rm cm^{-1}$ for a synthesized beam of $19.4'' \times 18.6''$.
    }
    \label{fivecontour}
\end{figure*}

\section{HI emission of SQ}
Figure 1  shows the contour map of the integrated HI emission between 5600 and 7000 km s$^{-1}$ overlaid on the  CFHT MegaCam r-band image. The large red circle in the center, which has a diameter of $12'$ (300 kpc in linear scale), represents the aperture within which we measure the SQ's HI spectrum. In Figure 2, the spectrum is compared with those obtained previously by mapping observations using the 305-meter Arecibo telescope \citep{1980ApJ...241L...1P} and the Green Bank Telescope (GBT; \citealt{Borthakur2015}). Thanks to its high sensitivity, the FAST spectrum is much less noisy than the previous two results.  From the FAST spectrum, the integrated HI flux of SQ is $20.5\pm 2.1$ Jy km s$^{-1}$. This is 1.2 times higher than result of Arecibo observations ($16.7 \pm 1.6$ Jy km s$^{-1}$) and 1.7 times higher than that of the GBT (11.8 Jy km s$^{-1}$) . The HI mass can be estimated using the formula $\rm M_{HI} = 2.356 \times 10^5 \, D^2 \, S \Delta \it v \rm \, M_\odot$, where  D is the comoving distance in Mpc and $\rm S\Delta \it v$ the integrated HI flux in Jy km s$^{-1}$). Accordingly, the FAST result corresponds to a HI mass of  $\rm M_{HI} = 3.48 \pm 0.35 \times 10^{10}\; M_\odot$ for SQ. 

With reference to previous high angular resolution synthetic observations by \citet{Shostak1984} and \citet{Williams2002}, which resolved the HI emission in SQ into several spatially and kinematically separated features, we divide the SQ spectrum into five velocity components: 5700 component (5600 --5800 km s$^{-1}$), 6000 component (5800 -- 6200 km s$^{-1}$), 6300 component (6200 -- 6400 km s$^{-1}$), 6600 component (6400 -- 6800 km s$^{-1}$), and 6900 component (6800 -- 7000 km s$^{-1}$. Note that these components do not include the emission outside the velocity range of 5600 -- 7000  km s$^{-1}$ whose contribution to the total HI flux is negligible (see Figure 2). The 5700 component corresponds to the ``West A'' feature in Westerbork Synthesis Radio Telescope HI map of \citet{Shostak1984} and the ``SW'' feature in VLA map of \citet{Williams2002}, the 6000 component to the ``West B'' feature in \citet{Shostak1984} and the ``NW-LV'' feature in \citet{Williams2002}, and the 6600 component to the ``Main'' feature in \citet{Shostak1984} and the combination of the ``NW-HV'', ``Arc-S'', and ``Arc-N'' features in \citet{Williams2002}. The 6300 and 6900 components are undetected in the two synthetic maps. The contour maps of the integrated emission (i.e. moment-0 maps) in the velocity ranges of  the five components are presented in the five panels of Figure \ref{fivecontour}. Identified in these maps are also several neighboring galaxies of SQ which will be discussed in detail in the next section. For comparison, contours of the high-resolution VLA maps adopted from \citet{Williams2002} are plotted in the figures for the 5700, 6000, and 6600 components.

As shown in Figures 3a and 3b, the 5700 and 6000 components are only marginally resolved by FAST. The 2-D Gaussian fittings of the unsmoothed FAST maps find sizes of 3$'$.2 and 3$'$.1 for the two components, respectively. These correspond to intrinsic sizes of 1$'$.5 and 1$'$.3 after the PSF deconvolution, consistent with the VLA measurements \citep{Williams2002}. According to  \citet{Williams2002}, the 5700 component is associated with the ``new intruder'' NGC 7318b whose radial velocity is 5774 km s$^{-1}$. Its HI mass measured by FAST is $\rm M_{HI} = 2.43\pm 0.24\; \times 10^9 \; M_\odot$. This is consistent with the result of Westerbork Synthesis Radio Telescope ($\rm M_{HI} = 2.4\pm 0.6\; \times 10^9 \; M_\odot$; \citealt{Shostak1984}) and higher than the VLA  value ($\rm 1.5\pm 0.2\;  \times  10^9 \; M_\odot$; \citealt{Williams2002}). \citet{1980ApJ...241L...1P} found a higher HI mass for this component in their Arecibo mapping observations ($\rm M_{HI}= 5.1\pm1.8\; \times 10^9 \; M_\odot$), although the difference from our result is only at 1.5$\sigma$ level because of the large error.

The origin of the 6000 component is still controversial. Based on the observations of \citet{Shostak1984}, \citet{Moles1997} suggested that it is also associated with NGC 7318b and the redshift difference from the 5700 component is due to the rotation of an HI disk. This was disputed by  \citet{Williams2002} who argued that the VLA observations, which have better angular resolution and flux sensitivity than those of the Westerbork Synthesis Radio Telescope, demonstrate that the 5700 and the 6000 components are distinctly separated both in the space and in the velocity domain, so are very unlikely to be associated with the same HI disk. We find an HI mass of ($\rm M_{HI} = 8.34\pm 0.83 \;\times 10^9\; M_\odot$) for the 6000 component. This is higher than the results in the literature, which are $\rm 5.3\pm 1.5\;  \times  10^9 \; M_\odot$, $\rm 3.1\pm 0.4\;  \times  10^9 \; M_\odot$ and $\rm 2.2\pm 0.3\;  \times  10^9 \; M_\odot$ by \citet{1980ApJ...241L...1P}, \citet{Shostak1984}, and \citet{Williams2002}, respectively.

Previously, the 6300 component of the HI 21 cm line emission was only
marginally detected by GBT observations \citep{Borthakur2010, Borthakur2015}. On the other hand, emissions in the velocity range
of 6200 -- 6400 km $s^{-1}$ have been detected in SQ in many other
bands, including the millimeter CO rotation line emission of molecular
gas \citep{Guillard2012, Yttergren2021}, the FIR [CII] and [OI] fine
structure line emissions \citep{Appleton2013, Appleton2017}, emissions
of the Balmer recombination lines and other emission lines of ionized
gas in the optical \citep{Xu2003, Iglesias-Paramo2012, DuartePuertas2019, Yttergren2021, Guillard2022}, and the Ly$\alpha$ line emission in UV \citep{Guillard2022}. All of these
detections of different emission lines, with angular resolutions
ranging from sub-arcsec to 50 arcsec, are consistent with the shock
excitation. The radiations are mainly from two locations: (1) the large shock front, (2) the so called ``bridge'' that links the Seyfert 2 nucleus of NGC 7319 and the shock front \citep{Cluver2010}. Some sub-regions in the large shock front show emissions in the velocity range of 6200 -- 6400 km s$^{-1}$ in the wings of relatively broad lines ($\Delta v > 300\; \rm  km\; s^{-1}$) peaked near 6000 or 6600 km s$^{-1}$ \citep{Guillard2012, Iglesias-Paramo2012, Appleton2013, Appleton2017, Guillard2022}. \citet{Guillard2012} attributed these emissions to the post-shock gas condensed out of the shock-heated hot X-ray gas after the cooling-down. There is a major difference between the large shock front and the bridge: in the radio continuum the former is very bright but the latter is undetected \citep{Xu2003}, indicating that the shocks in the two regions are different. The fact that the bridge is spatially and kinematically linked to the 4-kpc outflow \citep{Aoki1996} strongly suggests that the shocks in the bridge are powered by the outflow, and the radio quietness may be explained by the low efficiency of these shocks in accelerating cosmic-ray electrons. On the other hand, new JWST and ALMA observations with sub-arcsec resolutions detected in the bridge several filaments whose connection to the outflow is unclear \citep{Appleton2023}. The 2-D Gaussian fitting of the 6300 component finds a size of $4.0' \times 3.4'$, corresponding roughly to an intrinsic size of $\rm 2.9' \times 1.9'$ after the beam-deconvolution under the assumption that the emission can be approximated by a Gaussian source. Its HI mass is $\rm M_{HI} = 2.94\pm 0.30 \times 10^9\; M_\odot$.

The 6600 component, containing the majority of the HI emission associated with SQ (Figure \ref{totalHIspec}),  shows a strong central source plus low-level emissions in the outskirt. In particular,  an arm/tail-like diffuse feature on the south of SQ has been discovered within this component \citep{Xu2022}. The HI gas in the diffuse feature, which is outside the aperture within which the SQ spectrum is extracted (Figure 1), is confined in the velocity range of 6550 - 6750 km s$^{-1}$. The HI mass of the 6600 component of SQ is $\rm M_{HI} = 1.91\pm 0.19 \times 10^{10}\; M_\odot$, which is 56\% of the total HI mass of the SQ system.

The 6900 component was not detected in previous HI observations, but appeared in the observations of the CO emission of the molecular gas \citep{Guillard2012, Guillard2022} and in the surveys of HII regions \citep{Sulentic2001, Iglesias-Paramo2012, DuartePuertas2019}. The FAST map of the component (Figure \ref{fivecontour}e) centers very close to the intragroup starburst SQ-A \citep{Xu1999}, in agreement with the locations of the CO and H$\alpha$ emissions. The FAST observations do not resolve the component. The H$\alpha$ datacube of \citet{DuartePuertas2019}, which has an average angular resolution of $0.8''$, shows that this component is extended from SQ-A toward the shock front, presumably associated with pre-shock gas in the debris field. The HI mass of the 6900 component is $\rm M_{HI} = 1.38 \pm 0.14 \times 10^9\; M_\odot$.

The integrated fluxes and the corresponding HI masses of the five components are listed in Table 1. The errors include both the r.m.s and the calibration uncertainties (10\%; \citealt{Jiang2020}). The latter is the dominant contributor of the error in all cases.

\begin{figure*}
    \centering
    \includegraphics[width = 0.97\textwidth]{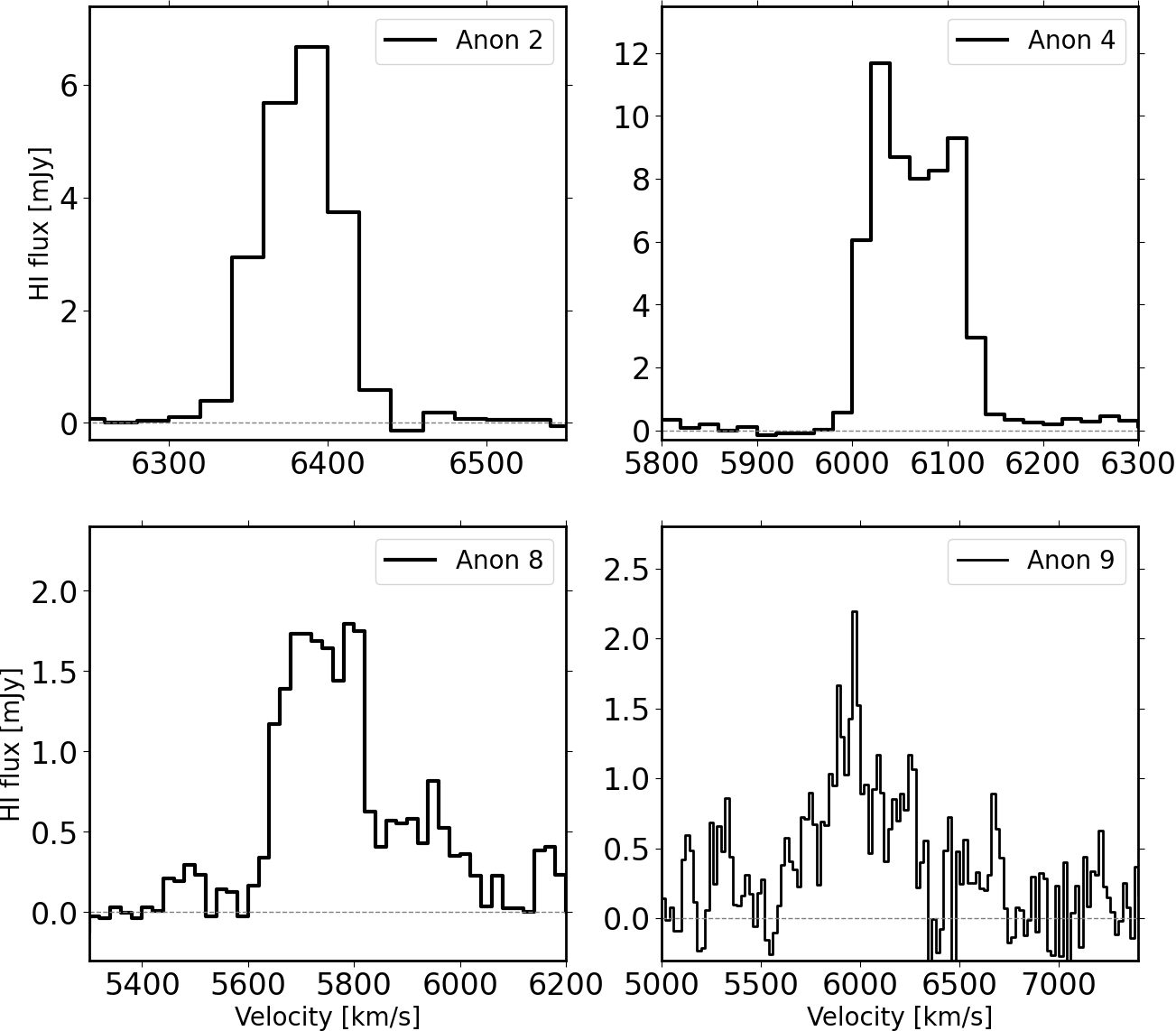}
    \caption{HI spectra of four neighboring galaxies. The dash gray lines show the position of zero flux. HI spectra of NGC~7320a and Anon~7 are presented in \citet{Xu2022}.
    }
    \label{SQsatellites}
\end{figure*}

\section{Neighboring galaxies in SQ field} \label{section: neighbors}

HI emission of six neighboring galaxies are detected in the SQ field. They are identified in Figure \ref{totalHImap}, Figure \ref{fivecontour} and listed in Table 2. These are low mass late-type galaxies with radial velocities in the range of 5700 -- 6800 km s$^{-1}$.  Among them, Anon 8 and Anon 9, named following the convention in the literature \citep{Shostak1984, Williams2002}, are new detections of this work. Anon 2 and Anon 4 were detected in previous observations \citep{Shostak1980, Williams2002}, and the FAST detections of NGC 7320a and Anon 7 were reported by \citet{Xu2022}. For Anon 2 and Anon 4, the HI velocities measured by FAST are consistent with those by VLA within 1-$\sigma$, and the values of $M_{\rm HI}$ by FAST are 0.1 dex and 0.2 dex above the VLA measurements, respectively, after the adjustments for the different $\rm H_0$ adopted (\citet{Williams2002} assumed $\rm H_0 = 75\; km\; s^{-1}\; Mpc^{-1}$ whereas we assume $\rm H_0 = 70\; km\; s^{-1}\; Mpc^{-1}$). 

Listed in Table 2 are also the optical counterparts of these neighboring galaxies, which are identified on the SDSS images with two criteria: (1) offset $< 1.5'$ and (2) $r < 20$ mag (corresponding roughly to $\rm M_* > 10^7 M_\odot$). In order to confirm their correspondence to the HI detections, we carried out long slit spectroscopic observations using the Lijiang 2.4 meter telescope for these optical sources (Appendix B). Five sources, including NGC 7320a reported by \citet{Xu2022}, are detected with their optical redshifts consistent with the HI velocities. Only the counterpart of Anon 7 is undetected spectroscopically. Anon 7 is very close to the bright background radio source B2 2233+33 (R.A. = 22h35m17.7s, Decl.=34d07m49s), with an offset of only 0.4'. Given the very high continuum flux density of B2 2233+33A ($\rm f_{1.4GHz} = 159.4 \pm 4.8\; mJy$; \citealt{Condon1998}) which is about 100 times of the peak HI flux density of Anon 7, we cannot rule out the possibility that Anon 7 is a pseudo detection due to artifacts associated with  B2 2233+33A. 
The optical redshifts and the stellar masses of the confirmed optical counterparts are presented in Table 2. The stellar masses are derived using the SDSS photometric data, the optical redshifts, and the code {\sc fastpp}\footnote{\url{https://github.com/cschreib/fastpp}} \citep{2009ApJ...700..221K}. The Chabrier IMF \citep{2003PASP..115..763C} and the stellar population model of \citet{BC03} are adopted. 

\section{Discussion}

\subsection{HI deficiency of compact groups}
It has been well established that spiral galaxies in compact groups of galaxies are deficient in HI gas content \citep[see the reviews in ][]{VerdesMontenegro2001, 2023A&A...670A..21J}. The VLA study of Hickson compact groups (HCGs) of \citet{VerdesMontenegro2001} show that many of them have a substantial amount of HI gas in the intragroup medium outside the member galaxies, and it is most likely that the intragroup HI is the stripped gas from the spiral galaxies in the groups. However, \citet{VerdesMontenegro2001} found that the HCGs are still HI deficient even after including the intragroup HI, with a mean HI deficiency $\rm Def_{HI} = 0.40\pm 0.07$.  The HI deficiency is defined by the equation $\rm Def_{HI} = \log (M_{HI, predicted}) - \log (M_{HI, observed})$, where  $\rm M_{HI, predicted}$ is the sum of the predicted HI mass of individual spiral galaxies in a group and the prediction is based on the optical luminosity and the morphology. \citet{VerdesMontenegro2001} proposed that the HI deficiency of HCGs is mainly due to the phase transition of the atomic gas in the intragroup medium during the evolution of compact groups. The GBT survey of HCGs by \citet{Borthakur2010} found a lower HI deficiency (with a mean  $\rm Def_{HI} = 0.20$), due both to the higher surface brightness sensitivity of GBT and the recovery of the missing fluxes in the VLA observations. SQ (also known by the name of HCG 92), as a prototype of HCGs, has all of its spiral members free of HI gas in their disks and its HI exclusively detected in the intragroup medium \citep{Shostak1980, Williams2002}. \citet{VerdesMontenegro2001} found a $\rm Def_{HI} = 0.49$ for SQ, which was reduced to  a $\rm Def_{HI} = 0.28$ by  \citet{Borthakur2010}. In this work, we found 1.7 times more HI than the  GBT observations of \citet{Borthakur2015} and 2.4 times of that of \citet{Borthakur2010}, thus SQ is not HI deficient anymore. The difference between our results and those of  \citet{Borthakur2015} cannot be explained by the calibration uncertainty which is at 10\% level for both FAST and GBT observations. A careful inspection of Figure \ref{totalHIspec}, where we compare the FAST spectrum with that of  \citet{Borthakur2015}, reveals that our FAST observations detected significantly more HI in the 6300 and 6900 component, namely in the wings on the two sides of the 6600 km s$^{-1}$ peak. Because of the lower sensitivity of GBT observations, these contributions (mainly due to low column density diffuse HI) were missed by  \citet{Borthakur2015}. In line with the results of \citet{Borthakur2010}, our FAST results further demonstrate that most of (or even all of) the missing HI in compact groups is in the low-density diffuse gas. Namely, instead of transforming to ionized gas, the low-density diffuse HI is persistent as neutral gas in the intragroup medium during the group evolution. It is worth noting that the diffuse HI gas discussed here is different from the  low-density HI gas in the diffuse feature discussed by  \citet{Xu2022}: the former is measured within the central aperture in Figure 1 and  the latter is outside the aperture. \citet{Xu2022} found that the mass of the HI gas in the diffuse feature is only $\sim 3\%$ of the HI mass of SQ and is therefore too low to affect the HI deficiency index of SQ significantly.

\begin{figure}
    \centering
    \includegraphics[width = 0.98\textwidth]{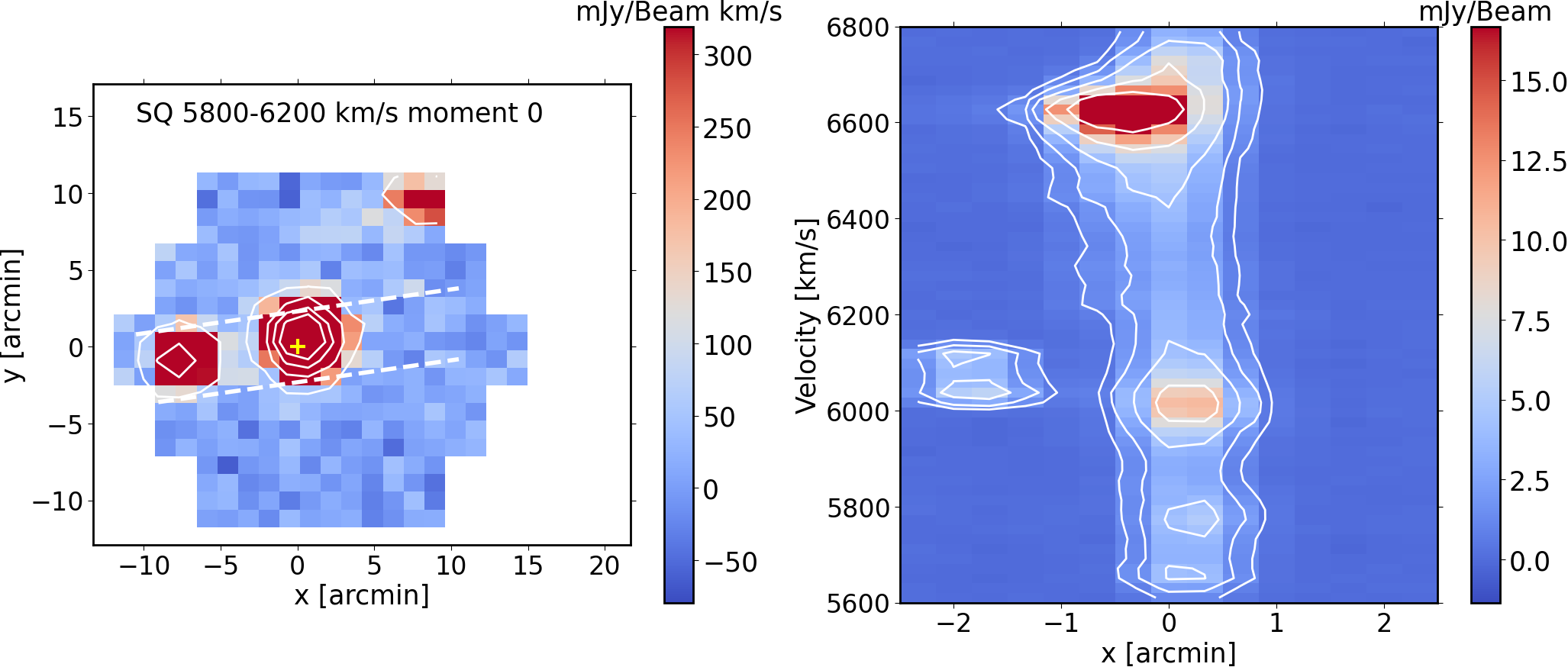}
    \caption{{\bf Left:} The contour and false color map of the 6000 component. The dashed white lines delineate the stripe within which the data of the P-V plot (in the right panel) are taken. The yellow cross marks the sky position of $x = 0$ in the P-V plot, whose coordinates are 22h36m00.8s +33d57m47s. The width of the stripe is 5' and the position angle is P.A. = 99 deg. {\bf Right:} The P-V plot. The contours start from 1 mJy beam$^{-1}$ (beam-width = 2.9') with an increment of a factor of 2. 
    }
    \label{pvplot}
\end{figure}

\subsection{Anon 4 and 6000 component}
One of the possible scenarios proposed by \citet{Xu2022} for the formation of the large diffuse feature in the 6600 component (Figure \ref{fivecontour}) is that, like the large Leo Ring \citep{Michel-Dansac2010}, the diffuse feature could be the product of a high-speed head-on collision between an unknown old intruder and one of the core members of SQ (i.e. NGC 7319, NGC 7318a, and NGC 7317). In this scenario, the collision triggers an expanding density wave that pushes gas in an extended HI disk of the target galaxy outwards to form a very large ring, of which the diffuse feature is the high-density part. A candidate for such an intruder could be Anon 4 ($v =$ 6057 km s$^{-1}$), which has a relative radial velocity of about 600 km s$^{-1}$ and a projected distance of approximately 0.2 Mpc from the SQ centre. If the relative transverse velocity is about 200 km s$^{-1}$, it would have taken approximately 1 Gyr for Anon 4 to move to the current position after the collision. If this scenario is true then it is conceivable that the 6000 component of the HI emission in the debris field of SQ is due to stripped gas from Anon 4 when it passed through the group center. In order to check this hypothesis we make a P-V plot along the line connecting Avon 4 and the center of the 6000 component (Figure \ref{pvplot}). The plot shows that the 6000 component (the emission around the position of x = 0 and velocity = 6000 km s$^{-1}$) has about the same velocity as Anon 4 (the emission around x = -2 arcmin), but there is no connection between the two emissions. This does not confirm any relationship between the 6000 component and Anon 4. Meanwhile, in the PV plot, there is a spur of the 6600 component toward Anon 4. The lower density gas between 6200 and 6400 km s$^{-1}$ (i.e. the 6300 component) has a similar (albeit weaker) trend. Compared to its counterparts in the 6000 and 6600 components, the central source of the 6300 component (Figure 3c) is slightly stretched toward the east, making its contours slightly elongated in the east-west direction (Figure \ref{fivecontour}). These trends might be suggestive of a tidal influence of Anon 4 on the HI gas in the debris field.

\subsection{Long history of interactions in SQ}
\citet{Moles1997} proposed a comprehensive picture for the interaction history of SQ based on a `two-intruders' scenario \citep[also see ][]{Sulentic2001}: an old intruder (NGC 7320c) stripped most of the gas from group members, and a new intruder (NGC 7318b) is currently colliding with this gas and triggering the large scale shock. The `old intruder' passed the core of SQ twice, one in $\sim 5$ - $7\;\times 10^8$ yr ago and pulled out the outer tail (also known as `old tail'), another about $2\;\times 10^8$ yr ago and triggered the inner tail (`young tail') . \citet{Xu2005} , based on newer observations, argued that it is more likely that the inner tail is triggered by NGC7318a which has a closer projected distance to NGC7319 than NGC7320c.  However, the simulations of \citet{Renaud2010} and \citet{Hwang2012} found that, if indeed the inner tail is generated later in a different encounter, it is very difficult for the outer tail to survive the strong tidal influence during the second passage. Instead, \citet{Hwang2012} concluded that it is most likely that the two tails are triggered by the same passage of NGC 7320c from a very close vicinity of NGC 7319 about  $9\times10^8$ years ago. The fiducial model of \citet{Hwang2012} can reasonably reproduce the the positions of the galaxies involved in the encounter between NGC 7319 and NGC 7320c and that between NGC 7318a and NGC 7318b, and the locations and the orientations of the inner and older tails. However, it cannot reproduce the overall HI distribution in SQ, in particular it predicts that NGC 7319 and NGC 7318b still retain most of their gas, in contradiction with the observations. For NGC 7318b, the fiducial model of \citet{Hwang2012} assumes that it has a relative radial velocity with NGC 7318a of $ \Delta v \sim 300\; \rm km\; s^{-1}$, which is significantly lower than the observed value ($\Delta v \sim 900\; \rm km\; s^{-1}$). This may cause an under-estimate of the strength of density waves triggered by the head-collision and result in less gas removal in NGC 7318b. For  NGC 7319, the assumption of the initial gas distribution as in a normal disk is unrealistic. As argued by \citet{Xu2022}, the diffuse HI feature attached to the south edge of the debris field was most likely produced by an interaction between a core galaxy of SQ and an earlier intruder (e.g. NGC 7320a or Anon 4) prior to the invasion of NGC 7320c. If the involved core galaxy were NGC 7319, then its gas disk had been already distorted when NGC 7320c arrived. Alternatively, NGC 7319 could have had been involved in a collision with the elliptical galaxy NGC7317 which seems to be linked with NGC 7319 with a diffuse optical halo \citep{Moles1997, Duc2018}. The recently published JWST MIRI images of SQ in the JWST Early Release Observations \citep{Pontoppidan2022} show some peculiar features in NGC 7319 in the 7.7 $\mu m$ band in incredible details, which is presumably dominated by the emission of PAH molecules. These features don't look like any tidal arms/tails at all, but are very similar to the spokes in a ring galaxy which are triggered by a head-on collision with an intruder galaxy (e.g. those in the Cartwheel Galaxy; \citealt{Renaud2018}).  The rather high molecular gas content and very low SFR in the disk of NGC 7319 \citep{Gao2000, Yttergren2021, Xu2005} correspond to a very low star formation efficiency, similar to the situation in Arp~142, a ring galaxy involved in a head-on collisions previously \citep{XU2021}. On the other hand, it is not clear whether the spoke-like features seen in JWST images were indeed produced by a head-on collision  $\sim 1$ Gyr ago, because according to the current theory the effects of such collisions shall fade away in a few $10^8$ yrs \citep{Renaud2018}. And even if they were, could they survive the tidal force during the subsequent passage of NGC 7320c? These question can only be answered in future simulations. 

\subsection{Origin of the diffuse HI feature}
Given its spatial and kinematical connections to the debris field produced by interactions among core members of SQ and by collisions with numerous intruders, \citet{Xu2022} argued that the diffuse feature was most likely due to gas ejected from one of the core galaxies during an interaction with an early intruder (plausibly NGC 7320a or Anon 4) more than 1 Gyr ago. \citet{Xu2022} ruled out the hypothesis that the diffuse feature is merely a collection of unresolved `dark' or `almost dark' galaxies \citep{Leisman2021} on the ground that it is very extended ($\sim 0.5$~Mpc) and has a narrow velocity range ($\Delta v_{20} = 160$~\rm km~s$^{-1}$). However, there are other possibilities.

First of all, it is possible that the diffuse feature is a tidal tail pull out from the neighboring spiral galaxy NGC~7320a instead of from a core member of SQ. The diffuse HI feature is right next to  NGC~7320a, and its radial velocity ($ v=6633~\rm km~s^{-1}$) is very close to that of NGC~7320a ($v=6702~\rm km~s^{-1}$). The simulations of \citet{Duc2008} suggested that high-speed encounters (relative velocity of $\sim 1000$ km s$^{-1}$) between two galaxies of nearly equal mass can produce long HI streams similar to the diffuse feature. Indeed there is an early type galaxy, KAZ~295, at the other end of the diffuse feature opposite to NGC~7320a. Our new optical spectroscopy finds a redshift velocity of 6555~km~s$^{-1}$ for the galaxy (see Appendix B), confirming it to be at about the same distance of NGC~7320a. Its stellar mass estimated from the SDSS photometry is $\rm M_* = 10^{9.71} M_\sun$ which is slightly lower than that of NGC~7320a ($\rm M_* = 10^{10.08} M_\sun$). In analog to the simulation of \citet{Duc2008}, KAZ~295 may have a high transverse velocity relative to NGC 7320a, and the encounter between the two may have pulled a long HI stream from NGC 7320a that is identified by us as the diffuse feature. However there are some inconsistencies which cast doubts on the analogy. The case simulated by \citet{Duc2008} are for 2 massive galaxies of $\rm M_* > 10^{11} M_\sun $, and NGC~7320a and KAZ~295 are low mass galaxies of $\rm M_* \sim 10^{10} M_\sun$. Also the simulated HI stream is only $\sim 200$ kpc whereas the length of the diffuse feature is $\sim 500$ kpc. Whether the putative interaction between the two low-mass galaxies can trigger the formation of an HI feature as long as 500 kpc needs to be investigated by future studies.

Another possibility is that the diffuse feature might be due to cold gas accretion by the SQ group from the intergalactic medium (IGM) in the large-scale filaments. The diffuse feature locates at the bottom of the FAST map and may well reach beyond the map. In the large-scale environment, SQ is in a rather high density region embedded within one of the main filaments in Perseus supercluster \citep{Gregory2000}. \citet{Tammann1970} suggested that SQ is perhaps part of the Zwicky Cluster $\rm 22h31m.2+37^\circ 32'$ \citep{ZwickyKowal1968}. That explains why there are many neighboring galaxies in the SQ field in a relatively large velocity range (5600 --7000 km s$^{-1}$), and why SQ has been intruded many times in its past history.

Smooth accretion of intergalactic gas from the surrounding filaments is a major growth mode for galaxy clusters and groups \citep{Walker2019}.  The cold intergalactic gas, which is mostly ionized and is $\sim 15\%$ of the intergalactic medium (IGM) in the filaments \citep{Cantalupo2012}, has been detected by observations of the Lyman-$\alpha$ emission and absorption \citep{Cantalupo2014, Finley2014, Umehata2019}, and by studies on the cross-correlations between neutral hydrogen (HI) 21 cm intensity maps and galaxy surveys \citep{Chang2010, Masui2013}. On the other hand, there is no confirmed detection of the HI 21 cm line emission of the filaments in the literature. Although the ``HI bridge'' detected in the intergalactic space between M31 and M33, which has a characteristic column density of $\rm N_{HI} \sim 10^{18}~cm^{-2}$, was first interpreted as the product of condensation within an intergalactic filament \citep{Braun2004, Lockman2012, Wolfe2016}, later simulations by \citet{Tepper-Garcia2022} suggests that it is more likely a tidal feature triggered by a close encounter between the two galaxies $\sim 6.5$ Gyr ago. Indeed, theoretical models predicted very low  neutral HI column density of the filaments, with a mean of $\rm N_{HI} \sim 10^{14}\; cm^{-2}$ for the total and a mean of $\rm N_{HI} \sim 10^{16}\; cm^{-2}$ for the brightest 10\% of the filaments  \citep{Kooistra2019}. This is significantly lower than the characteristic HI column density of the diffuse feature ($\rm N_{HI} \sim 7 \times 10^{17} \; cm^{-2}$;  \citealt{Xu2022}). Nevertheless, this does not completely reject the accretion hypothesis for the origin of the diffuse feature. Future HI observations exploring the regions beyond the south boundary of our FAST map, preferably with even higher sensitivity, will provide crucial constraints to this hypothesis.

\section{Summary}
HI mapping observations were carried out using the FAST 19-beam receiver. The observations cover a sky region of size of $\sim 30'\times 30'$ around SQ with an angular resolution of $2.9'$ and a sampling that satisfies the Nyquist criterion. The smoothed sensitivity of the data (resolution = $4'$) reaches $\rm 1\sigma = 4.3\times 10^{16}\; cm^{-2}$  for $20\; \rm km\; s^{-1}$ channels. The discovery of a large HI structure of radial velocity of $ 6633~\rm km~s^{-1}$ was presented in \citet{Xu2022}. In this paper. we report the following new results:

\begin{itemize}
\item The HI spectrum of SQ covers a wide velocity range between 5600 and 7000 km s$^{-1}$, with an integrated flux of $20.5\pm 2.1$ Jy km s$^{-1}$. The corresponding HI mass is  $\rm M_{HI} = 3.48 \pm 0.35 \times 10^{10}\; M_\sun$ (for a comoving distance of 85 Mpc). This is significantly higher than the previous measurements  by the VLA and GBT observations \citep{VerdesMontenegro2001, Williams2002, Borthakur2010, Borthakur2015} and indicates that, when including the HI gas in the intragroup medium, SQ is not HI deficient.

\item The HI emission of SQ can be divided into five kinematical components: 5700 component (5600 -- 5800 km s$^{-1}$), 6000 component (6000 -- 6200 km s$^{-1}$), 6300 component (6200 -- 6400 km s$^{-1}$), 6600 component (6400 -- 6800 km s$^{-1}$), and 6900 component (6800 -- 7000 km s$^{-1}$). Whereas the 5700, 6000 and 6600 components have been detected previously, the 6300 and 6900 components were undetected or only marginally detected in previous observations due to inadequate sensitivity. The net HI mass of the  6300 and 6900 components is $\rm 5.71 \pm 0.42 \times 10^{19}\; M_\sun$, which is $\rm 16.4\pm 1.6\%$ of the total HI of SQ.

\item The HI emission of six neighboring galaxies are detected in the SQ field. These include NGC 7320a, Anon 2, Anon 4, Anon 7, Anon 8 and Anon 9. These are low mass late-type galaxies with radial velocities in the range of 5700 -- 6800 km s$^{-1}$. Among them, Anon 2 and Anon 4 were detected in previous observations \citep{Shostak1980, Williams2002}, and the FAST detections of NGC 7320a and Anon 7 were reported by \citet{Xu2022}. We obtain the optical redshifts for optical counterparts of Anon 2, Anon 4, Anon 8 and Anon 9 which are consistent with the HI radial velocities.

\item The FAST observations provide new evidence for a complex and long history of SQ, particularly before the formation of the inner and outer tails. The newly discovered diffuse feature is most likely due to a gas ejection from one of the core galaxies during an interaction with an early intruder (NGC 7320a or Anon 4) more than 1 Gyr ago. However the possibilities that the diffuse feature is an HI stream pull out from the neighboring low-mass spiral NGC~7320a by a high-speed encounter with another low-mass galaxy (KAZ~295), and that the diffuse feature might be due to cold gas accretion by the SQ group from the intergalactic medium in the large-scale filaments, warrant further investigations.

\end{itemize}

\appendix

\section{HI channel maps}
The Figure A1 presents the channel maps of the FAST observation, covering the velocity range between 5500 and 7080 km s$^{-1}$ with 80 channels of width $ \Delta v = 20\rm km\; s^{-1}$. In each channel map, contours of the HI emission (with smoothed angular resolution of 4') are overlaid on the inverted grey-scale map of the deep CFHT MegaCam r-band image \citep{Duc2018}. The contours start from $\rm N_{HI} = 1.3 \times 10^{17}\; cm^{-2}$ (at 3$\sigma$ level) with the increment of a factor of 2. The central source associated with the SQ proper is detected in most channels. The diffuse feature reported in \citet{Xu2022} is visible close to the bottom of the field in channels centered at $v = $ 6620, 6640, and 6660 km s$^{-1}$. Several neighboring galaxies are also detected in different channels across the velocity range, which are discussed in details in section \ref{section: neighbors}.

\section{Optical spectroscopic observations of neighboring galaxies}
We performed long slit spectroscopic observations for the optical counterparts of ANON~2, ANON~4, ANON~7, ANON~8 and ANON~9, and for the optical galaxy KAZ~295, using Lijiang 2.4m telescope on Oct. 30-31 and Dec. 14, 2022. A slit of $1.8''$ wide was chosen which matches the seeing.  We used the G14 grating with a wavelength coverage from 3500 - 7000 with a resolution about 1000, which can help us to cover the bright emission lines such as H$\alpha$, [OIII], as well as the [OII], Ca II H, K lines at bluer wavelengths. The exposure time is about 2 hours for each target. We reduced the long slit spectra by the standard IRAF script. The HeNe lamp was used for wavelength calibration. However, bright lines of the lamp spectra were saturated, which affected our wavelength accuracy. Thus, conservatively, we included a wavelength calibration error of $1 \sigma = 3$\AA\ (137 km/s). Since the spectra are for redshift measurements only, no flux calibration was performed. The spectra of the optical counterparts of Anon~2, Anon~4, and Anon~8, and of the optical galaxy KAZ~295, are presented in Figure \ref{optspecB1}. The spectrum of the optical counterpart of Anon~9, which is a weak detection, is presented in Figure \ref{optspecB2}. The units of the flux density ($F_\lambda$) in the figures are arbitrary.


\begin{deluxetable*}{lcccccccccc}
\tabletypesize{\normalsize}
\tablewidth{0pt}
\tablecaption{HI kinematical components of SQ.}
\label{HIdynam}
\tablehead{
Name                          &  Velocity range    & Integrated flux\rlap{\tablenotemark{a}} & $\rm M_{HI}$ \\
                              & km s$^{-1}$ & Jy km s$^{-1}$  & $\rm \times 10^9 M_\odot$}
\startdata
5700                & 5600 - 5800 &  1.46 $\pm$ 0.15 &  2.49 $\pm$ 0.25 \\
6000                & 5800 - 6200 &  5.01 $\pm$ 0.50 &  8.52 $\pm$ 0.85 \\
6300                & 6200 - 6400 &  1.76 $\pm$ 0.18 &  2.99 $\pm$ 0.30 \\
6600                & 6400 - 6800 & 11.42 $\pm$ 1.14 & 19.45 $\pm$ 1.95 \\
6900                & 6800 - 7000 &  0.83 $\pm$ 0.08 &  1.37 $\pm$ 0.14 \\
total               & 5600 - 7000 & 20.46 $\pm$ 2.05 & 34.86 $\pm$ 3.49 \\
\enddata
\tablenotetext{a}{
Errors include the rms and the calibration uncertainty (10\%).
}
\end{deluxetable*}


\begin{deluxetable*}{lcccccccccc}
\tabletypesize{\scriptsize}
\tablewidth{0pt}
\tablecaption{Neighboring galaxies detected by FAST observations}
\label{HItargets}
\tablehead{
ID  & R.A.    & Decl.    & HI Velocity  & log($M_{\rm HI}/M_{\odot}$) & optical ID    & log($M_*/M_{\odot}$)\tablenotemark{a} & optical velocity\tablenotemark{b}  & references\tablenotemark{c}    \\
    & J2000 & J2000  & km s$^{-1}$  & $\rm log(M_\odot)$          &               & $\rm log(M_\odot)$    &  km s$^{-1}$                   & 
    }
\startdata
Anon 2                     & 22:35:50.7 & +34:09:30  & 6391 $\pm$ 8  & 8.90 $\pm$ 0.04   & SDSSJ223550.70+340930.8  & 8.58  & 6325 $\pm$ 140  & 1, 2  \\
Anon 4                     & 22:36:40.9 & +33:56:54  & 6075 $\pm$ 13 & 9.30 $\pm$ 0.04   & SDSSJ223640.85+335654.5  & 8.62  & 6089 $\pm$ 141 & 1, 2  \\
Anon 7\tablenotemark{d}  & 22:35:19.3 & +34:08:00  & 6654 $\pm$ 16 & 8.34 $\pm$ 0.10   & SDSSJ223526.26+340837.9  &   &                     & 3      \\
Anon 8                     & 22:35:47.5 & +34:06:48  & 5789 $\pm$ 20 & 8.88 $\pm$ 0.04   & SDSSJ223547.54+340648.1  & 9.46  & 5625 $\pm$ 151 & 1    \\
Anon 9  & 22:35:22.7 & +34:08:03  & 5976 $\pm$ 38 & 9.01 $\pm$ 0.05   & SDSSJ223526.26+340837.9  & 8.12  & 5886 $\pm$ 145 & 1   \\
NGC 7320a\tablenotemark{e} & 22:36:32.2 & +33:47:46  & 6702 $\pm$ 24 & 8.80 $\pm$ 0.04   & NGC 7320a               & 10.08  & 6729 $\pm$ 59  & 3 \\
\enddata
\tablenotetext{a}{Stellar mass derived from SDSS photometry and optical redshift.}
\tablenotetext{b}{Optical velocity in the local standard of rest  (LSR) reference frame.}
\tablenotetext{c}{References: 1. This work; 2. \citet{Williams2002}; 3. \citet{Xu2022} }
\tablenotetext{d}{HI data taken from \citet{Xu2022}; uncertain optical identification.}
\tablenotetext{e}{HI data and optical velocity taken from \citet{Xu2022}.}
\end{deluxetable*}

\begin{acknowledgments}
We thank the referee for careful reading and constructive suggestions.
  This work is supported by the National Key R\&D Programme of China No. 2017YFA0402704 and National Natural Science Foundation of China (NSFC) No. 11873055 and sponsored (in part) by the Chinese Academy of Sciences (CAS) through a grant to the CAS South America Center for Astronomy. C.K.X. acknowledges NSFC grant No. 11733006. C.C. acknowledges NSFC grant No. 11803044 and 12173045. We acknowledge the science research grants from the China Manned Space Project with NO. CMS-CSST-2021-A05.
  N.-Y.T. is supported by the National key R\&D program of China under grant no. 2018YFE0202900 and the Cultivation Project for FAST Scientific Payoff and Research Achievement of CAMS-CAS. Y.S. Dai acknowledges support from the National Key R\&D Program of China for grant No. \ 2022YFA1605300 and the National Nature Science Foundation of China (NSFC) grants No.\ 12273051 and \ 11933003; J.-S.H. acknowledges NSFC grant No. 11933003. U.L. acknowledges support from project PID2020-114414GB-100, financed by MCIN/AEI/10.13039/501100011033, from project P20\_00334 financed by the Junta de Andaluc\'ia and from FEDER/Junta de Andaluc\'ia-Consejer\'ia de Transformaci\'{o}n Econ\'{o}mica, Industria, Conocimiento y Universidades/Proyecto A-FQM-510-UGR20. F.R. acknowledges support from the Knut and Alice Wallenberg Foundation. This work made use of data from FAST, a Chinese national mega-science facility built and operated by the National Astronomical Observatories, CAS. We thank P. Jiang, L. Hou, C. Sun and other FAST operation team members for supports in the observations and data reductions, and Y. Huang, K.-X. Lu for helping with the optical spectroscopic observations. 
  C.C. appreciates helpful comments about the usage of matplotlib from Dr. Wenda Zhang. Support of the staff from the Lijiang 2.4 m telescope is acknowledged. Funding for the Lijiang 2.4 m telescope has been provided by the CAS and the People’s Government of Yunnan Province. This research has made use of the NASA/IPAC Extragalactic Database, which is operated by the Jet Propulsion Laboratory, California Institute of Technology, under contract with the National Aeronautics and Space Administration.
\end{acknowledgments}

%

\vspace{5mm}
\facilities{FAST (19-Beam Receiver), Lijiang 2m4 telescope (YFOSC)}


\software{astropy \citep{2013A&A...558A..33A,2018AJ....156..123A},  
          Source Extractor \citep{1996A&AS..117..393B},
          pvextractor\citep{2016ascl.soft08010G},
          Radio Astronomy Tools in Python\citep{2015ASPC..499..363G}
          }

\restartappendixnumbering 

\setcounter{figure}{0}
\renewcommand{\thefigure}{A\arabic{figure}}
\begin{figure*}
    \centering
    \includegraphics[width=0.18\textwidth]{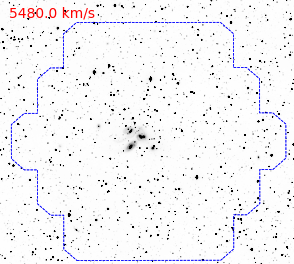}
    \includegraphics[width=0.18\textwidth]{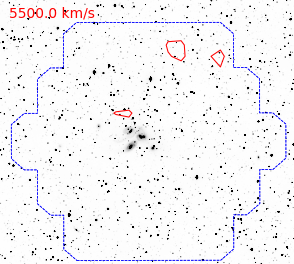}
    \includegraphics[width=0.18\textwidth]{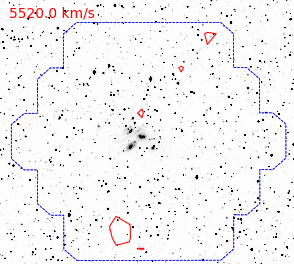}
    \includegraphics[width=0.18\textwidth]{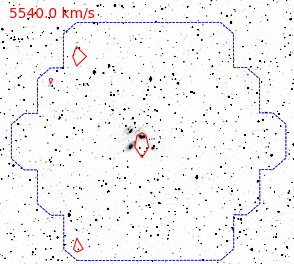}
    \includegraphics[width=0.18\textwidth]{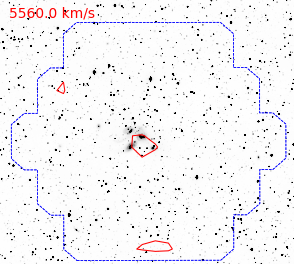}
    \includegraphics[width=0.18\textwidth]{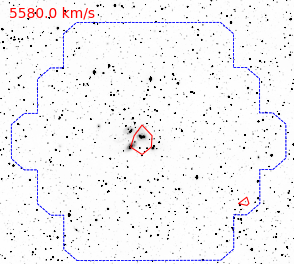}
    \includegraphics[width=0.18\textwidth]{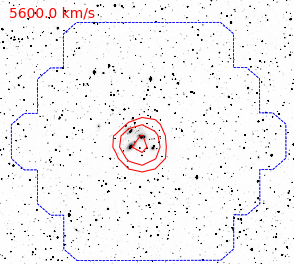}
    \includegraphics[width=0.18\textwidth]{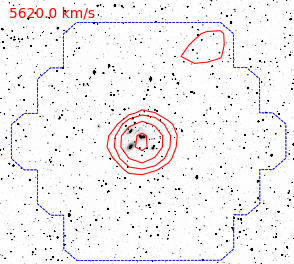}
    \includegraphics[width=0.18\textwidth]{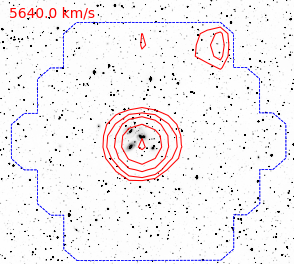}
    \includegraphics[width=0.18\textwidth]{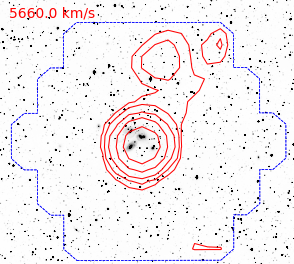}
    \includegraphics[width=0.18\textwidth]{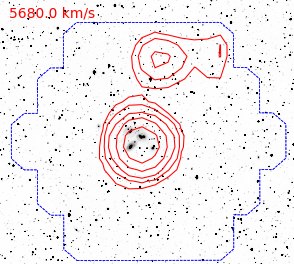}
    \includegraphics[width=0.18\textwidth]{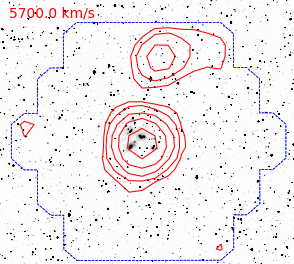}
    \includegraphics[width=0.18\textwidth]{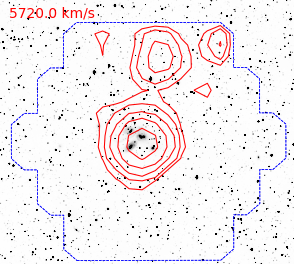}
    \includegraphics[width=0.18\textwidth]{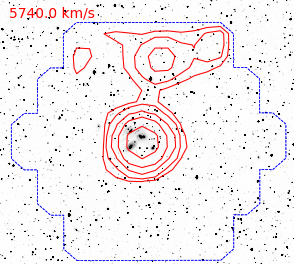}
    \includegraphics[width=0.18\textwidth]{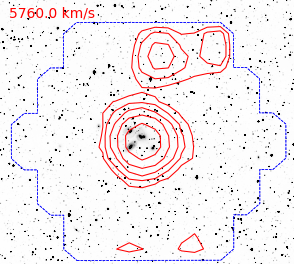}
    \includegraphics[width=0.18\textwidth]{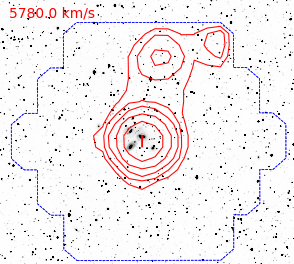}
    \includegraphics[width=0.18\textwidth]{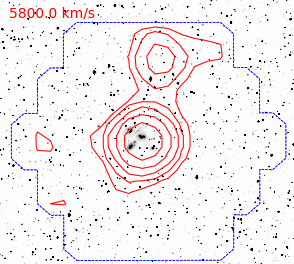}
    \includegraphics[width=0.18\textwidth]{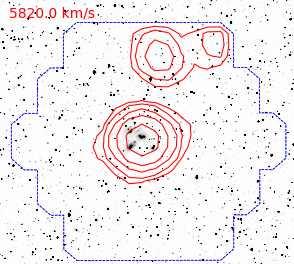}
    \includegraphics[width=0.18\textwidth]{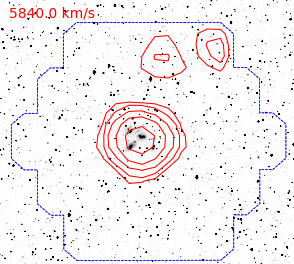}
    \includegraphics[width=0.18\textwidth]{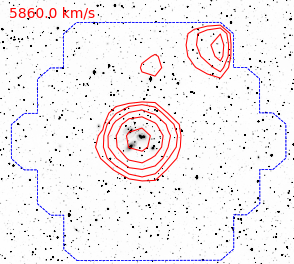}
    \includegraphics[width=0.18\textwidth]{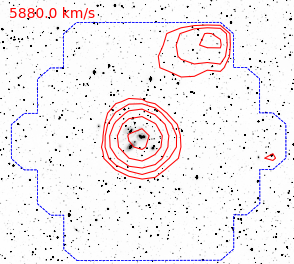}
    \includegraphics[width=0.18\textwidth]{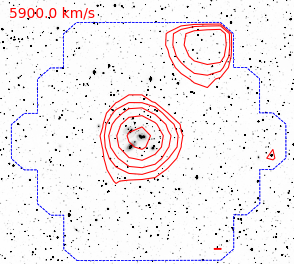}
    \includegraphics[width=0.18\textwidth]{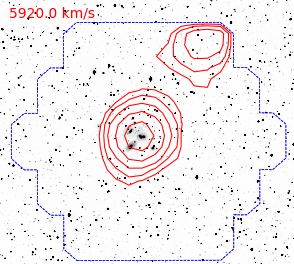}
    \includegraphics[width=0.18\textwidth]{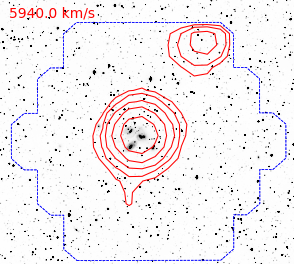}
    \includegraphics[width=0.18\textwidth]{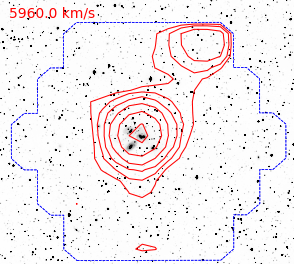}
    \includegraphics[width=0.18\textwidth]{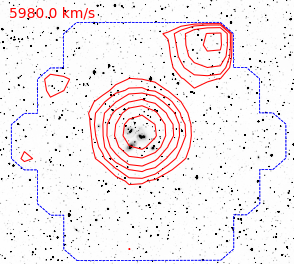}
    \includegraphics[width=0.18\textwidth]{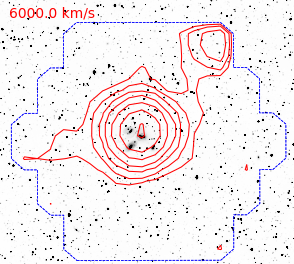}
    \includegraphics[width=0.18\textwidth]{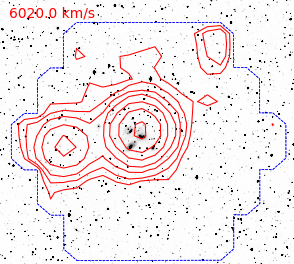}
    \includegraphics[width=0.18\textwidth]{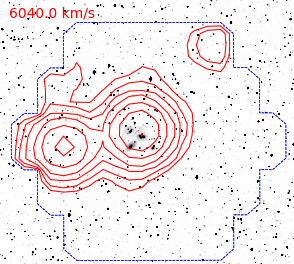}
    \includegraphics[width=0.18\textwidth]{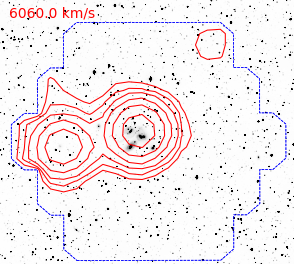}
    \includegraphics[width=0.18\textwidth]{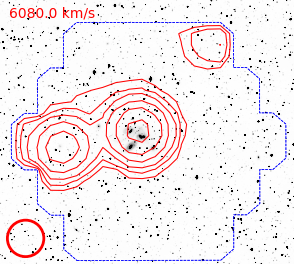}
    \includegraphics[width=0.18\textwidth]{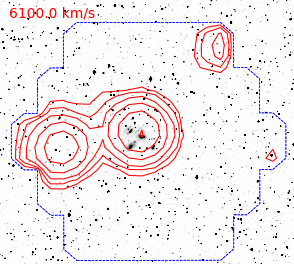}
    \includegraphics[width=0.18\textwidth]{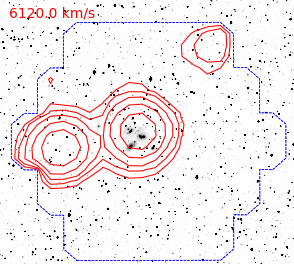}
    \includegraphics[width=0.18\textwidth]{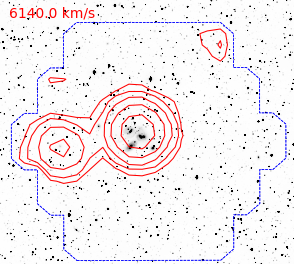}
    \includegraphics[width=0.18\textwidth]{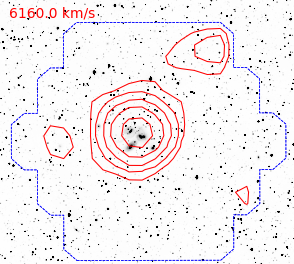}
    \caption{FAST HI channel maps (contours) over-plotted on the CFHT r-band image \citep{Duc2018}. The channel maps cover the velocity range of 5480 -- 7040 km s$^{-1}$ with a channel width of 20 km s$^{-1}$. The contour levels are [3, 6, 12, 24, 48, 96...]$\times 4.2 \times 10^{16}\; \rm cm^{-2}$. The blue boundaries delineate the FAST coverage. The small red circle at lower left corner of the first map in the last row  shows the angular resolution of the HI channel maps (FWHM = $4'$). 
    }
    \label{contour1}
\end{figure*}

\clearpage
\setcounter{figure}{0}
\begin{figure*}
    \centering
    \includegraphics[width=0.18\textwidth]{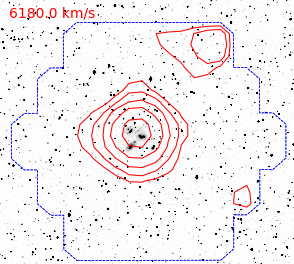}
    \includegraphics[width=0.18\textwidth]{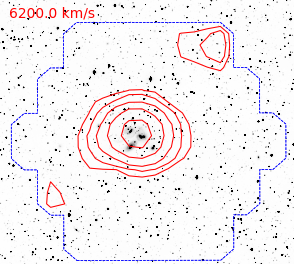}
    \includegraphics[width=0.18\textwidth]{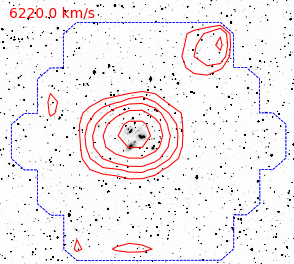}
    \includegraphics[width=0.18\textwidth]{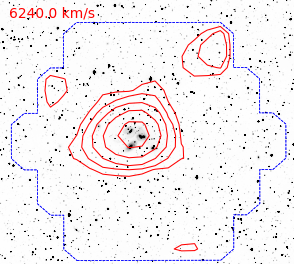}
    \includegraphics[width=0.18\textwidth]{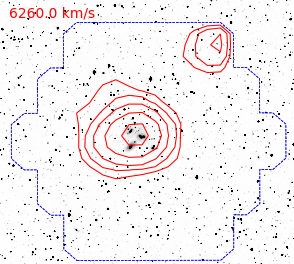}
    \includegraphics[width=0.18\textwidth]{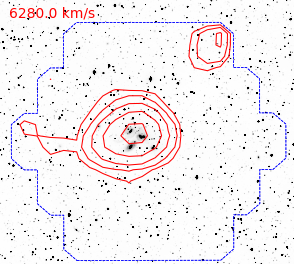}
    \includegraphics[width=0.18\textwidth]{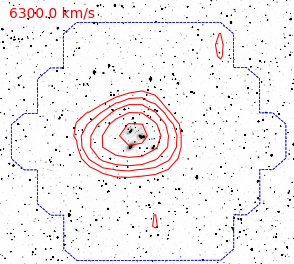}
    \includegraphics[width=0.18\textwidth]{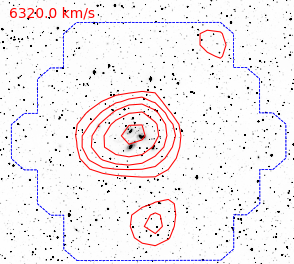}
    \includegraphics[width=0.18\textwidth]{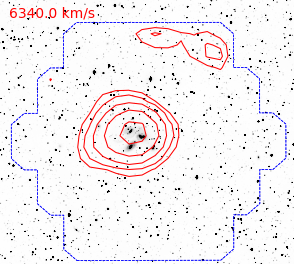}
    \includegraphics[width=0.18\textwidth]{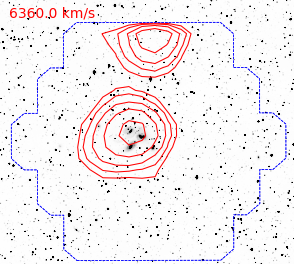}
    \includegraphics[width=0.18\textwidth]{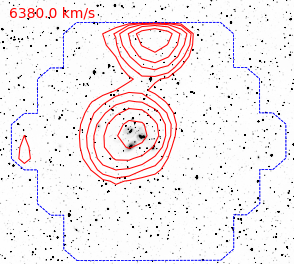}
    \includegraphics[width=0.18\textwidth]{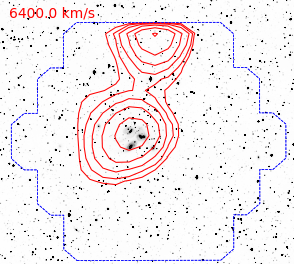}
    \includegraphics[width=0.18\textwidth]{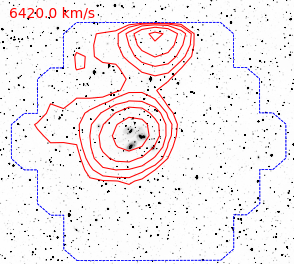}
    \includegraphics[width=0.18\textwidth]{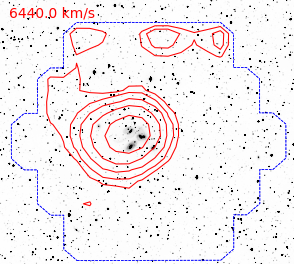}
    \includegraphics[width=0.18\textwidth]{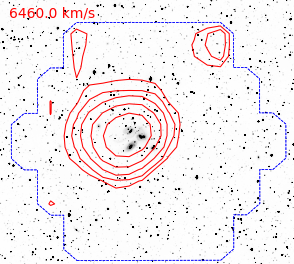}
    \includegraphics[width=0.18\textwidth]{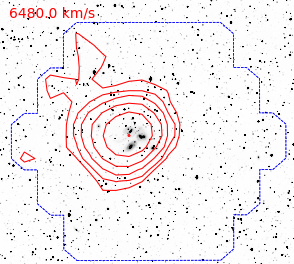}
    \includegraphics[width=0.18\textwidth]{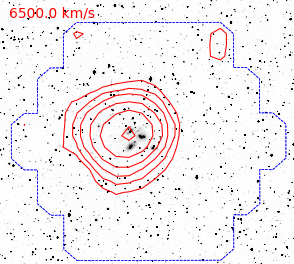}
    \includegraphics[width=0.18\textwidth]{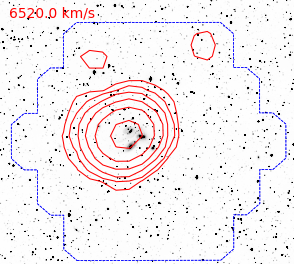}
    \includegraphics[width=0.18\textwidth]{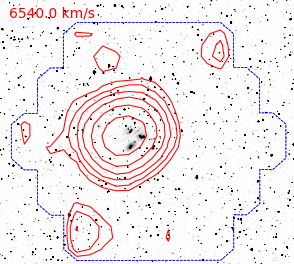}
    \includegraphics[width=0.18\textwidth]{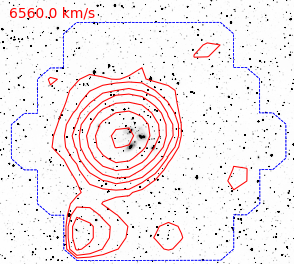}
    \includegraphics[width=0.18\textwidth]{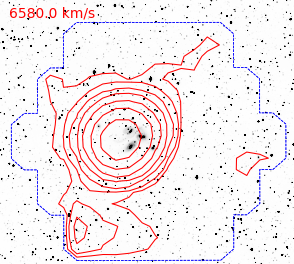}
    \includegraphics[width=0.18\textwidth]{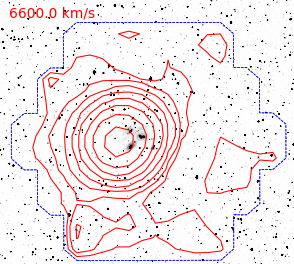}
    \includegraphics[width=0.18\textwidth]{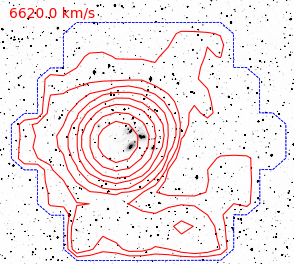}
    \includegraphics[width=0.18\textwidth]{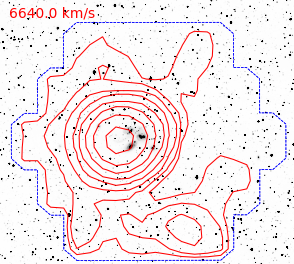}
    \includegraphics[width=0.18\textwidth]{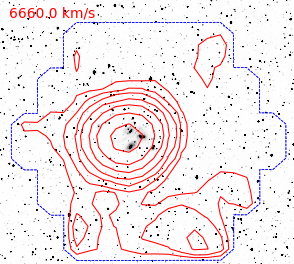}
    \includegraphics[width=0.18\textwidth]{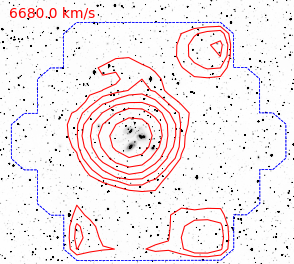}
    \includegraphics[width=0.18\textwidth]{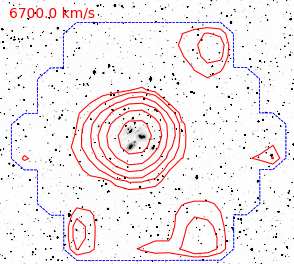}
    \includegraphics[width=0.18\textwidth]{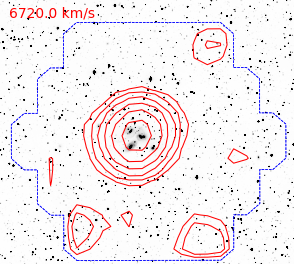}
    \includegraphics[width=0.18\textwidth]{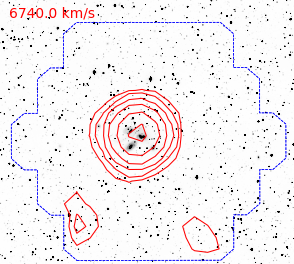}
    \includegraphics[width=0.18\textwidth]{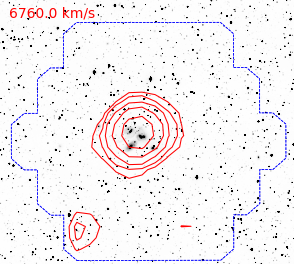}
    \includegraphics[width=0.18\textwidth]{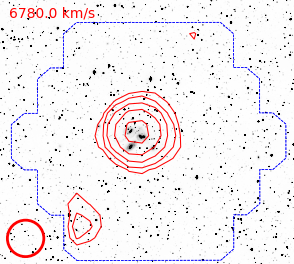}
    \includegraphics[width=0.18\textwidth]{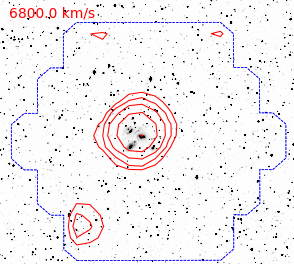}
    \includegraphics[width=0.18\textwidth]{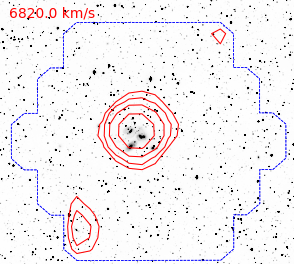}
    \includegraphics[width=0.18\textwidth]{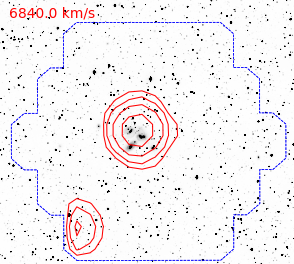}
    \includegraphics[width=0.18\textwidth]{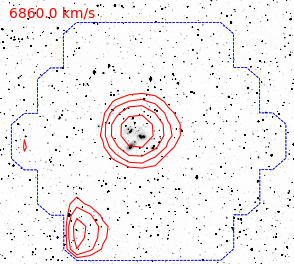}
    \caption{{\bf (Continued)} Contours of HI channel maps over-plotted on the r-band image.}
    \label{contour2}
\end{figure*}

\clearpage
\setcounter{figure}{0}
\begin{figure*}[ht!]
\centering
\includegraphics[width=0.18\textwidth]{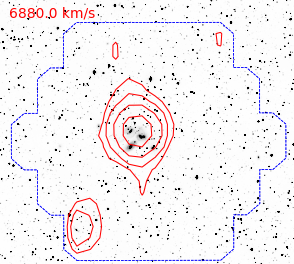}
\includegraphics[width=0.18\textwidth]{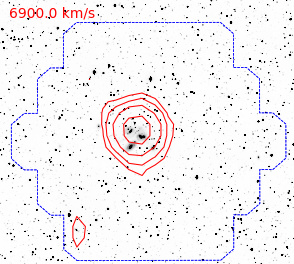}
\includegraphics[width=0.18\textwidth]{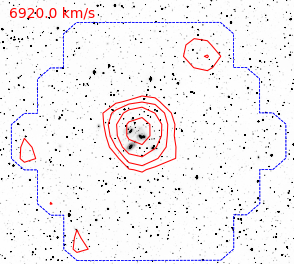}
\includegraphics[width=0.18\textwidth]{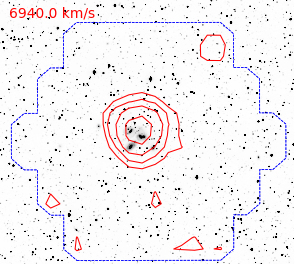}
\includegraphics[width=0.18\textwidth]{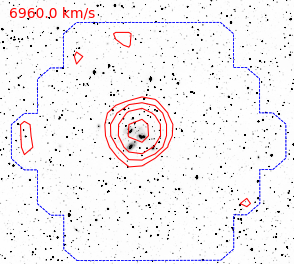}
\includegraphics[width=0.18\textwidth]{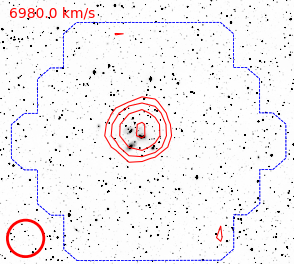}
\includegraphics[width=0.18\textwidth]{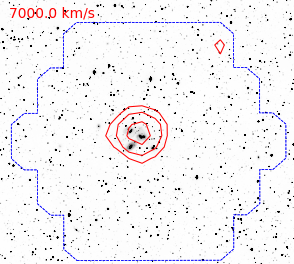}
\includegraphics[width=0.18\textwidth]{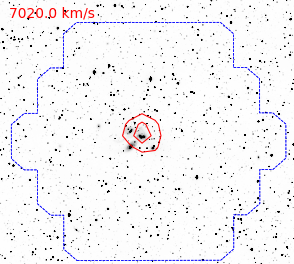}
\includegraphics[width=0.18\textwidth]{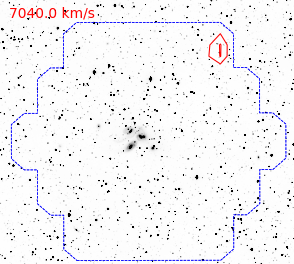}
\includegraphics[width=0.18\textwidth]{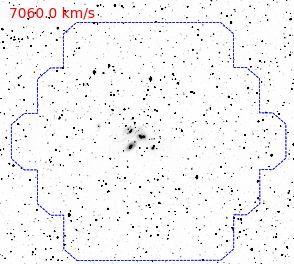}
 \caption{{\bf (Continued)} Contours of HI channel maps over-plotted on the r-band image.}  
    \label{contour3}
\end{figure*}

\setcounter{figure}{0}
\renewcommand{\thefigure}{B\arabic{figure}}

\begin{figure*}[ht!]
\centering
\includegraphics[width=0.9\textwidth]{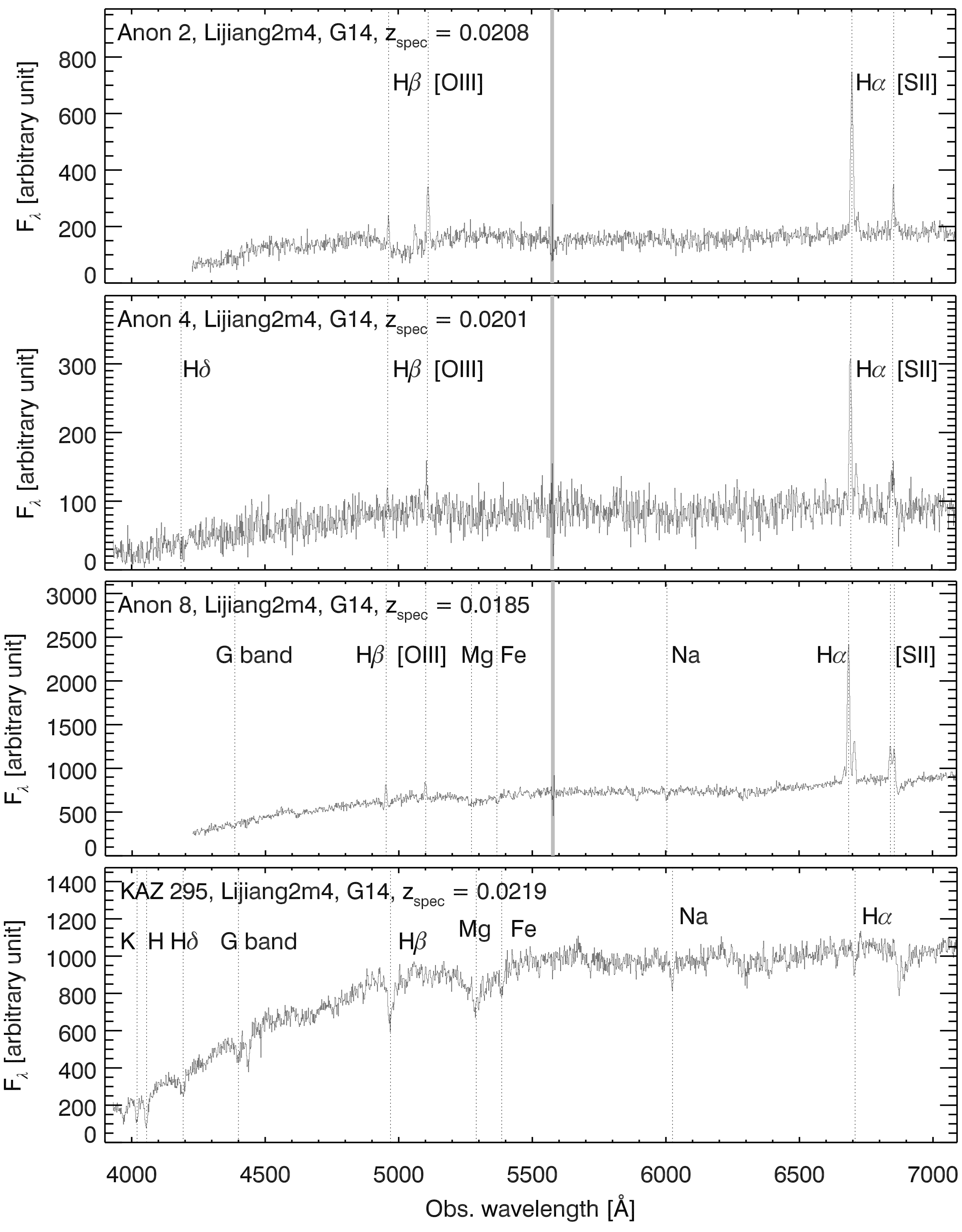}
 \caption{
 Optical spectra. The grey regions are contaminated by the sky lines. 
 }  
\label{optspecB1}
\end{figure*}

\begin{figure*}[ht!]
\centering
\includegraphics[width=0.98\textwidth]{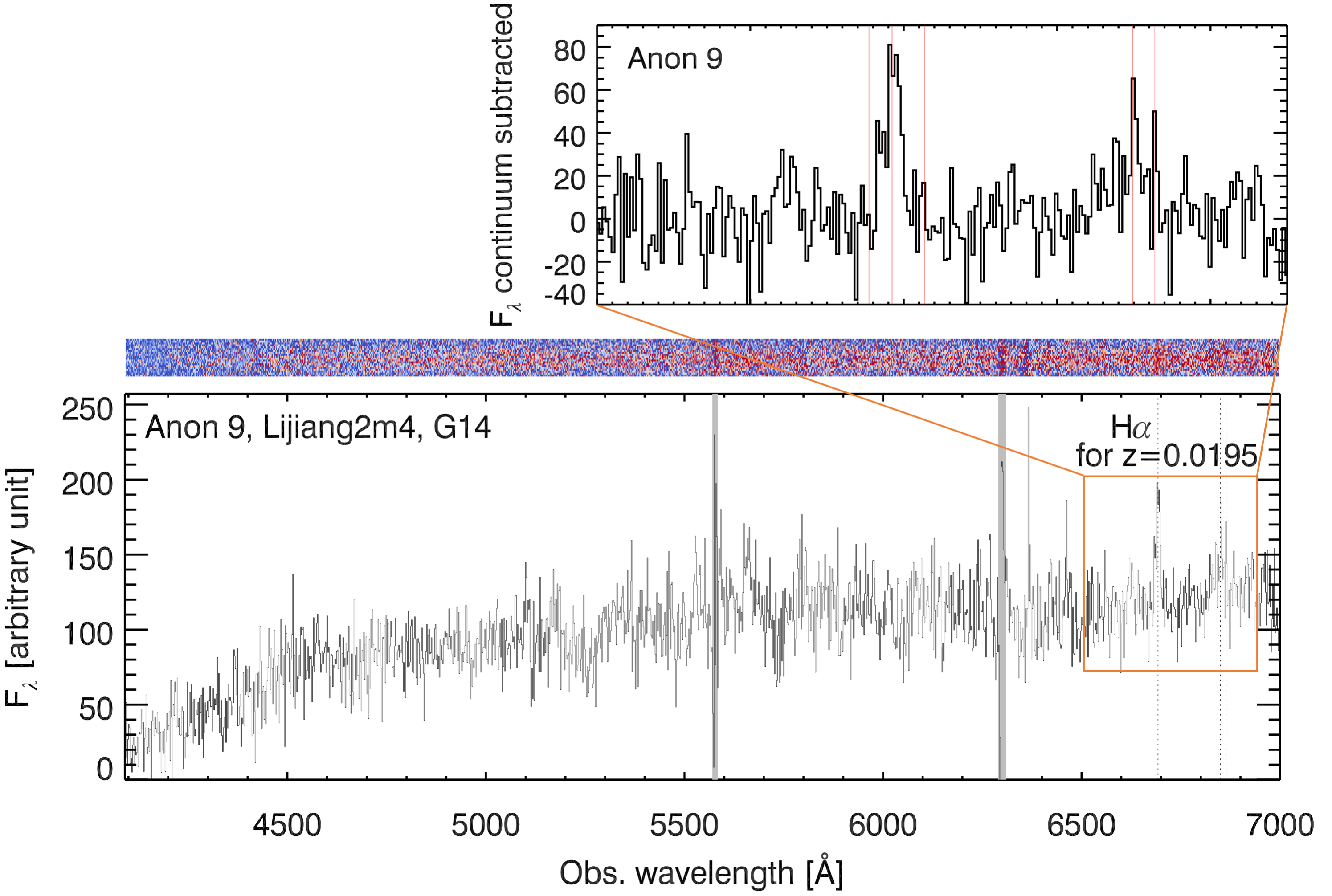}
\caption{
{\bf Top panel:} Continuum subtracted spectrum of the optical counterpart of Anon 9 between 6500\AA - 6950\AA observed wavelength. The red lines indicate the [NII], H$\alpha$ and [SII] position at redshift 0.0195. {\bf Middle panel: }2-D spectrum of the Anon 9 with the wavelength range the same as the lower panel. {\bf Lower panel:} 1-D original spectrum. The dash lines show the position of H$\alpha$ and [SII] at z = 0.0195. The gray regions are contaminated by the sky lines. 
 } 
\label{optspecB2}
\end{figure*}




\begin{thebibliography}{}
\expandafter\ifx\csname natexlab\endcsname\relax\def\natexlab#1{#1}\fi
\providecommand{\url}[1]{\href{#1}{#1}}
\providecommand{\dodoi}[1]{doi:~\href{http://doi.org/#1}{\nolinkurl{#1}}}
\providecommand{\doeprint}[1]{\href{http://ascl.net/#1}{\nolinkurl{http://ascl.net/#1}}}
\providecommand{\doarXiv}[1]{\href{https://arxiv.org/abs/#1}{\nolinkurl{https://arxiv.org/abs/#1}}}

\bibitem[{{Alatalo} {et~al.}(2014){Alatalo}, {Appleton}, {Lisenfeld},
  {Bitsakis}, {Guillard}, {Charmandaris}, {Cluver}, {Dopita}, {Freeland},
  {Jarrett}, {Kewley}, {Ogle}, {Rasmussen}, {Rich}, {Verdes-Montenegro}, {Xu},
  \& {Yun}}]{Alatalo2014}
{Alatalo}, K., {Appleton}, P.~N., {Lisenfeld}, U., {et~al.} 2014, \apj, 795,
  159, \dodoi{10.1088/0004-637X/795/2/159}

\bibitem[{{Allen} \& {Hartsuiker}(1972)}]{Allen1972}
{Allen}, R.~J., \& {Hartsuiker}, J.~W. 1972, \nat, 239, 324,
  \dodoi{10.1038/239324a0}

\bibitem[{{Aoki} {et~al.}(1996){Aoki}, {Ohtani}, {Yoshida}, \&
  {Kosugi}}]{Aoki1996}
{Aoki}, K., {Ohtani}, H., {Yoshida}, M., \& {Kosugi}, G. 1996, \aj, 111, 140,
  \dodoi{10.1086/117767}

\bibitem[{{Appleton} {et~al.}(2006){Appleton}, {Xu}, {Reach}, {Dopita}, {Gao},
  {Lu}, {Popescu}, {Sulentic}, {Tuffs}, \& {Yun}}]{Appleton2006}
{Appleton}, P.~N., {Xu}, K.~C., {Reach}, W., {et~al.} 2006, \apjl, 639, L51,
  \dodoi{10.1086/502646}

\bibitem[{{Appleton} {et~al.}(2013){Appleton}, {Guillard}, {Boulanger},
  {Cluver}, {Ogle}, {Falgarone}, {Pineau des For{\^e}ts}, {O'Sullivan}, {Duc},
  {Gallagher}, {Gao}, {Jarrett}, {Konstantopoulos}, {Lisenfeld}, {Lord}, {Lu},
  {Peterson}, {Struck}, {Sturm}, {Tuffs}, {Valchanov}, {van der Werf}, \&
  {Xu}}]{Appleton2013}
{Appleton}, P.~N., {Guillard}, P., {Boulanger}, F., {et~al.} 2013, \apj, 777,
  66, \dodoi{10.1088/0004-637X/777/1/66}

\bibitem[{{Appleton} {et~al.}(2017){Appleton}, {Guillard}, {Togi}, {Alatalo},
  {Boulanger}, {Cluver}, {Pineau des For{\^e}ts}, {Lisenfeld}, {Ogle}, \&
  {Xu}}]{Appleton2017}
{Appleton}, P.~N., {Guillard}, P., {Togi}, A., {et~al.} 2017, \apj, 836, 76,
  \dodoi{10.3847/1538-4357/836/1/76}

\bibitem[{{Appleton} {et~al.}(2023){Appleton}, {Guillard}, {Emonts},
  {Boulanger}, {Togi}, {Reach}, {Alatalo}, {Cluver}, {Diaz Santos}, {Duc},
  {Gallagher}, {Ogle}, {O'Sullivan}, {Voggel}, \& {Xu}}]{Appleton2023}
{Appleton}, P.~N., {Guillard}, P., {Emonts}, B., {et~al.} 2023, arXiv e-prints,
  arXiv:2301.02928, \dodoi{10.48550/arXiv.2301.02928}

\bibitem[{{Astropy Collaboration} {et~al.}(2013){Astropy Collaboration},
  {Robitaille}, {Tollerud}, {Greenfield}, {Droettboom}, {Bray}, {Aldcroft},
  {Davis}, {Ginsburg}, {Price-Whelan}, {Kerzendorf}, {Conley}, {Crighton},
  {Barbary}, {Muna}, {Ferguson}, {Grollier}, {Parikh}, {Nair}, {Unther},
  {Deil}, {Woillez}, {Conseil}, {Kramer}, {Turner}, {Singer}, {Fox}, {Weaver},
  {Zabalza}, {Edwards}, {Azalee Bostroem}, {Burke}, {Casey}, {Crawford},
  {Dencheva}, {Ely}, {Jenness}, {Labrie}, {Lim}, {Pierfederici}, {Pontzen},
  {Ptak}, {Refsdal}, {Servillat}, \& {Streicher}}]{2013A&A...558A..33A}
{Astropy Collaboration}, {Robitaille}, T.~P., {Tollerud}, E.~J., {et~al.} 2013,
  \aap, 558, A33, \dodoi{10.1051/0004-6361/201322068}

\bibitem[{{Astropy Collaboration} {et~al.}(2018){Astropy Collaboration},
  {Price-Whelan}, {Sip{\H{o}}cz}, {G{\"u}nther}, {Lim}, {Crawford}, {Conseil},
  {Shupe}, {Craig}, {Dencheva}, {Ginsburg}, {VanderPlas}, {Bradley},
  {P{\'e}rez-Su{\'a}rez}, {de Val-Borro}, {Aldcroft}, {Cruz}, {Robitaille},
  {Tollerud}, {Ardelean}, {Babej}, {Bach}, {Bachetti}, {Bakanov}, {Bamford},
  {Barentsen}, {Barmby}, {Baumbach}, {Berry}, {Biscani}, {Boquien}, {Bostroem},
  {Bouma}, {Brammer}, {Bray}, {Breytenbach}, {Buddelmeijer}, {Burke},
  {Calderone}, {Cano Rodr{\'\i}guez}, {Cara}, {Cardoso}, {Cheedella}, {Copin},
  {Corrales}, {Crichton}, {D'Avella}, {Deil}, {Depagne}, {Dietrich}, {Donath},
  {Droettboom}, {Earl}, {Erben}, {Fabbro}, {Ferreira}, {Finethy}, {Fox},
  {Garrison}, {Gibbons}, {Goldstein}, {Gommers}, {Greco}, {Greenfield},
  {Groener}, {Grollier}, {Hagen}, {Hirst}, {Homeier}, {Horton}, {Hosseinzadeh},
  {Hu}, {Hunkeler}, {Ivezi{\'c}}, {Jain}, {Jenness}, {Kanarek}, {Kendrew},
  {Kern}, {Kerzendorf}, {Khvalko}, {King}, {Kirkby}, {Kulkarni}, {Kumar},
  {Lee}, {Lenz}, {Littlefair}, {Ma}, {Macleod}, {Mastropietro}, {McCully},
  {Montagnac}, {Morris}, {Mueller}, {Mumford}, {Muna}, {Murphy}, {Nelson},
  {Nguyen}, {Ninan}, {N{\"o}the}, {Ogaz}, {Oh}, {Parejko}, {Parley}, {Pascual},
  {Patil}, {Patil}, {Plunkett}, {Prochaska}, {Rastogi}, {Reddy Janga},
  {Sabater}, {Sakurikar}, {Seifert}, {Sherbert}, {Sherwood-Taylor}, {Shih},
  {Sick}, {Silbiger}, {Singanamalla}, {Singer}, {Sladen}, {Sooley},
  {Sornarajah}, {Streicher}, {Teuben}, {Thomas}, {Tremblay}, {Turner},
  {Terr{\'o}n}, {van Kerkwijk}, {de la Vega}, {Watkins}, {Weaver}, {Whitmore},
  {Woillez}, {Zabalza}, \& {Astropy Contributors}}]{2018AJ....156..123A}
{Astropy Collaboration}, {Price-Whelan}, A.~M., {Sip{\H{o}}cz}, B.~M., {et~al.}
  2018, \aj, 156, 123, \dodoi{10.3847/1538-3881/aabc4f}

\bibitem[{{Bertin} \& {Arnouts}(1996)}]{1996A&AS..117..393B}
{Bertin}, E., \& {Arnouts}, S. 1996, \aaps, 117, 393,
  \dodoi{10.1051/aas:1996164}

\bibitem[{{Borthakur} {et~al.}(2010){Borthakur}, {Yun}, \&
  {Verdes-Montenegro}}]{Borthakur2010}
{Borthakur}, S., {Yun}, M.~S., \& {Verdes-Montenegro}, L. 2010, \apj, 710, 385,
  \dodoi{10.1088/0004-637X/710/1/385}

\bibitem[{{Borthakur} {et~al.}(2015){Borthakur}, {Yun}, {Verdes-Montenegro},
  {Heckman}, {Zhu}, \& {Braatz}}]{Borthakur2015}
{Borthakur}, S., {Yun}, M.~S., {Verdes-Montenegro}, L., {et~al.} 2015, \apj,
  812, 78, \dodoi{10.1088/0004-637X/812/1/78}

\bibitem[{{Braun}(2004)}]{Braun2004}
{Braun}, R. 2004, \nar, 48, 1271, \dodoi{10.1016/j.newar.2004.09.010}

\bibitem[{{Bruzual} \& {Charlot}(2003)}]{BC03}
{Bruzual}, G., \& {Charlot}, S. 2003, \mnras, 344, 1000,
  \dodoi{10.1046/j.1365-8711.2003.06897.x}

\bibitem[{{Cantalupo} {et~al.}(2014){Cantalupo}, {Arrigoni-Battaia},
  {Prochaska}, {Hennawi}, \& {Madau}}]{Cantalupo2014}
{Cantalupo}, S., {Arrigoni-Battaia}, F., {Prochaska}, J.~X., {Hennawi}, J.~F.,
  \& {Madau}, P. 2014, \nat, 506, 63, \dodoi{10.1038/nature12898}

\bibitem[{{Cantalupo} {et~al.}(2012){Cantalupo}, {Lilly}, \&
  {Haehnelt}}]{Cantalupo2012}
{Cantalupo}, S., {Lilly}, S.~J., \& {Haehnelt}, M.~G. 2012, \mnras, 425, 1992,
  \dodoi{10.1111/j.1365-2966.2012.21529.x}

\bibitem[{{Chabrier}(2003)}]{2003PASP..115..763C}
{Chabrier}, G. 2003, \pasp, 115, 763, \dodoi{10.1086/376392}

\bibitem[{{Chang} {et~al.}(2010){Chang}, {Pen}, {Bandura}, \&
  {Peterson}}]{Chang2010}
{Chang}, T.-C., {Pen}, U.-L., {Bandura}, K., \& {Peterson}, J.~B. 2010, \nat,
  466, 463, \dodoi{10.1038/nature09187}

\bibitem[{{Cheng} {et~al.}(2020){Cheng}, {Ibar}, {Du}, {Molina},
  {Orellana-Gonz{\'a}les}, {Zhang}, {Zhu}, {Xu}, {Wu}, {Cao}, {Huang},
  {Leiton}, {Hughes}, {He}, {Li}, {Xu}, {Dai}, {Shao}, \& {Musin}}]{Cheng2020}
{Cheng}, C., {Ibar}, E., {Du}, W., {et~al.} 2020, \aap, 638, L14,
  \dodoi{10.1051/0004-6361/202038483}

\bibitem[{{Cluver} {et~al.}(2010){Cluver}, {Appleton}, {Boulanger}, {Guillard},
  {Ogle}, {Duc}, {Lu}, {Rasmussen}, {Reach}, {Smith}, {Tuffs}, {Xu}, \&
  {Yun}}]{Cluver2010}
{Cluver}, M.~E., {Appleton}, P.~N., {Boulanger}, F., {et~al.} 2010, \apj, 710,
  248, \dodoi{10.1088/0004-637X/710/1/248}

\bibitem[{{Cluver} {et~al.}(2013){Cluver}, {Appleton}, {Ogle}, {Jarrett},
  {Rasmussen}, {Lisenfeld}, {Guillard}, {Verdes-Montenegro}, {Antonucci},
  {Bitsakis}, {Charmandaris}, {Boulanger}, {Egami}, {Xu}, \&
  {Yun}}]{Cluver2013}
{Cluver}, M.~E., {Appleton}, P.~N., {Ogle}, P., {et~al.} 2013, \apj, 765, 93,
  \dodoi{10.1088/0004-637X/765/2/93}

\bibitem[{{Condon} {et~al.}(1998){Condon}, {Cotton}, {Greisen}, {Yin},
  {Perley}, {Taylor}, \& {Broderick}}]{Condon1998}
{Condon}, J.~J., {Cotton}, W.~D., {Greisen}, E.~W., {et~al.} 1998, \aj, 115,
  1693, \dodoi{10.1086/300337}

\bibitem[{{Duarte Puertas} {et~al.}(2019){Duarte Puertas},
  {Iglesias-P{\'a}ramo}, {Vilchez}, {Drissen}, {Kehrig}, \&
  {Martin}}]{DuartePuertas2019}
{Duarte Puertas}, S., {Iglesias-P{\'a}ramo}, J., {Vilchez}, J.~M., {et~al.}
  2019, \aap, 629, A102, \dodoi{10.1051/0004-6361/201935686}

\bibitem[{{Duc} \& {Bournaud}(2008)}]{Duc2008}
{Duc}, P.-A., \& {Bournaud}, F. 2008, \apj, 673, 787, \dodoi{10.1086/524868}

\bibitem[{{Duc} {et~al.}(2018){Duc}, {Cuillandre}, \& {Renaud}}]{Duc2018}
{Duc}, P.-A., {Cuillandre}, J.-C., \& {Renaud}, F. 2018, \mnras, 475, L40,
  \dodoi{10.1093/mnrasl/sly004}

\bibitem[{{En{\ss}lin}(2002)}]{Ensslin2002}
{En{\ss}lin}, T.~A. 2002, \aap, 396, L17, \dodoi{10.1051/0004-6361:20021613}

\bibitem[{{Fedotov} {et~al.}(2011){Fedotov}, {Gallagher}, {Konstantopoulos},
  {Chandar}, {Bastian}, {Charlton}, {Whitmore}, \& {Trancho}}]{Fedotov2011}
{Fedotov}, K., {Gallagher}, S.~C., {Konstantopoulos}, I.~S., {et~al.} 2011,
  \aj, 142, 42, \dodoi{10.1088/0004-6256/142/2/42}

\bibitem[{{Finley} {et~al.}(2014){Finley}, {Petitjean}, {Noterdaeme}, \&
  {P{\^a}ris}}]{Finley2014}
{Finley}, H., {Petitjean}, P., {Noterdaeme}, P., \& {P{\^a}ris}, I. 2014, \aap,
  572, A31, \dodoi{10.1051/0004-6361/201423961}

\bibitem[{{Gallagher} {et~al.}(2001){Gallagher}, {Charlton}, {Hunsberger},
  {Zaritsky}, \& {Whitmore}}]{Gallagher2001}
{Gallagher}, S.~C., {Charlton}, J.~C., {Hunsberger}, S.~D., {Zaritsky}, D., \&
  {Whitmore}, B.~C. 2001, \aj, 122, 163, \dodoi{10.1086/321111}

\bibitem[{{Gao} \& {Xu}(2000)}]{Gao2000}
{Gao}, Y., \& {Xu}, C. 2000, \apjl, 542, L83, \dodoi{10.1086/312940}

\bibitem[{{Ginsburg} {et~al.}(2016){Ginsburg}, {Robitaille}, \&
  {Beaumont}}]{2016ascl.soft08010G}
{Ginsburg}, A., {Robitaille}, T., \& {Beaumont}, C. 2016, {pvextractor:
  Position-Velocity Diagram Extractor}.
\newblock \doeprint{1608.010}

\bibitem[{{Ginsburg} {et~al.}(2015){Ginsburg}, {Robitaille}, {Beaumont},
  {Rosolowsky}, {Leroy}, {Brogan}, {Hunter}, {Teuben}, \&
  {Brisbin}}]{2015ASPC..499..363G}
{Ginsburg}, A., {Robitaille}, T., {Beaumont}, C., {et~al.} 2015, in
  Astronomical Society of the Pacific Conference Series, Vol. 499, Revolution
  in Astronomy with ALMA: The Third Year, ed. D.~{Iono}, K.~{Tatematsu},
  A.~{Wootten}, \& L.~{Testi}, 363--364

\bibitem[{{Gregory} {et~al.}(2000){Gregory}, {Tifft}, {Moody}, {Newberry}, \&
  {Hall}}]{Gregory2000}
{Gregory}, S.~A., {Tifft}, W.~G., {Moody}, J.~W., {Newberry}, M.~V., \& {Hall},
  S.~M. 2000, \aj, 119, 567, \dodoi{10.1086/301196}

\bibitem[{{Guillard} {et~al.}(2012){Guillard}, {Boulanger}, {Pineau des
  For{\^e}ts}, {Falgarone}, {Gusdorf}, {Cluver}, {Appleton}, {Lisenfeld},
  {Duc}, {Ogle}, \& {Xu}}]{Guillard2012}
{Guillard}, P., {Boulanger}, F., {Pineau des For{\^e}ts}, G., {et~al.} 2012,
  \apj, 749, 158, \dodoi{10.1088/0004-637X/749/2/158}

\bibitem[{{Guillard} {et~al.}(2022){Guillard}, {Appleton}, {Boulanger},
  {Shull}, {Lehnert}, {Pineau des Forets}, {Falgarone}, {Cluver}, {Xu},
  {Gallagher}, \& {Duc}}]{Guillard2022}
{Guillard}, P., {Appleton}, P.~N., {Boulanger}, F., {et~al.} 2022, \apj, 925,
  63, \dodoi{10.3847/1538-4357/ac313f}

\bibitem[{{Hickson}(1982)}]{Hickson1982}
{Hickson}, P. 1982, \apj, 255, 382, \dodoi{10.1086/159838}

\bibitem[{{Hwang} {et~al.}(2012){Hwang}, {Struck}, {Renaud}, \&
  {Appleton}}]{Hwang2012}
{Hwang}, J.-S., {Struck}, C., {Renaud}, F., \& {Appleton}, P.~N. 2012, \mnras,
  419, 1780, \dodoi{10.1111/j.1365-2966.2011.19847.x}

\bibitem[{{Iglesias-P{\'a}ramo} {et~al.}(2012){Iglesias-P{\'a}ramo},
  {L{\'o}pez-Mart{\'\i}n}, {V{\'\i}lchez}, {Petropoulou}, \&
  {Sulentic}}]{Iglesias-Paramo2012}
{Iglesias-P{\'a}ramo}, J., {L{\'o}pez-Mart{\'\i}n}, L., {V{\'\i}lchez}, J.~M.,
  {Petropoulou}, V., \& {Sulentic}, J.~W. 2012, \aap, 539, A127,
  \dodoi{10.1051/0004-6361/201118055}

\bibitem[{{Jiang} {et~al.}(2019){Jiang}, {Yue}, {Gan}, {Yao}, {Li}, {Pan},
  {Sun}, {Yu}, {Liu}, {Tang}, {Qian}, {Lu}, {Yan}, {Peng}, {Zhang}, {Wang},
  {Li}, \& {Li}}]{2019SCPMA..6259502J}
{Jiang}, P., {Yue}, Y., {Gan}, H., {et~al.} 2019, Science China Physics,
  Mechanics, and Astronomy, 62, 959502, \dodoi{10.1007/s11433-018-9376-1}

\bibitem[{{Jiang} {et~al.}(2020){Jiang}, {Tang}, {Hou}, {Liu}, {Kr{\v{c}}o},
  {Qian}, {Sun}, {Ching}, {Liu}, {Duan}, {Yue}, {Gan}, {Yao}, {Li}, {Pan},
  {Yu}, {Liu}, {Li}, {Peng}, {Yan}, \& {FAST Collaboration}}]{Jiang2020}
{Jiang}, P., {Tang}, N.-Y., {Hou}, L.-G., {et~al.} 2020, Research in Astronomy
  and Astrophysics, 20, 064, \dodoi{10.1088/1674-4527/20/5/64}

\bibitem[{{Johnson} {et~al.}(2007){Johnson}, {Hibbard}, {Gallagher},
  {Charlton}, {Hornschemeier}, {Jarrett}, \& {Reines}}]{Johnson2007}
{Johnson}, K.~E., {Hibbard}, J.~E., {Gallagher}, S.~C., {et~al.} 2007, \aj,
  134, 1522, \dodoi{10.1086/520921}

\bibitem[{{Jones} {et~al.}(2023){Jones}, {Verdes-Montenegro}, {Moldon}, {Damas
  Segovia}, {Borthakur}, {Luna}, {Yun}, {del Olmo}, {Perea}, {Cannon}, {Lopez
  Gutierrez}, {Cluver}, {Garrido}, \& {Sanchez}}]{2023A&A...670A..21J}
{Jones}, M.~G., {Verdes-Montenegro}, L., {Moldon}, J., {et~al.} 2023, \aap,
  670, A21, \dodoi{10.1051/0004-6361/202244622}

\bibitem[{{Kooistra} {et~al.}(2019){Kooistra}, {Silva}, {Zaroubi}, {Verheijen},
  {Tempel}, \& {Hess}}]{Kooistra2019}
{Kooistra}, R., {Silva}, M.~B., {Zaroubi}, S., {et~al.} 2019, \mnras, 490,
  1415, \dodoi{10.1093/mnras/stz2677}

\bibitem[{{Kriek} {et~al.}(2009){Kriek}, {van Dokkum}, {Labb{\'e}}, {Franx},
  {Illingworth}, {Marchesini}, \& {Quadri}}]{2009ApJ...700..221K}
{Kriek}, M., {van Dokkum}, P.~G., {Labb{\'e}}, I., {et~al.} 2009, \apj, 700,
  221, \dodoi{10.1088/0004-637X/700/1/221}

\bibitem[{{Leisman} {et~al.}(2021){Leisman}, {Rhode}, {Ball}, {Pagel},
  {Cannon}, {Salzer}, {Janowiecki}, {Janesh}, {J{\'o}zsa}, {Giovanelli},
  {Haynes}, {Adams}, {Gray}, \& {Smith}}]{Leisman2021}
{Leisman}, L., {Rhode}, K.~L., {Ball}, C., {et~al.} 2021, \aj, 162, 274,
  \dodoi{10.3847/1538-3881/ac2a38}

\bibitem[{{Lenki{\'c}} {et~al.}(2016){Lenki{\'c}}, {Tzanavaris}, {Gallagher},
  {Desjardins}, {Walker}, {Johnson}, {Fedotov}, {Charlton}, {Hornschemeier},
  {Durrell}, \& {Gronwall}}]{Lenkic2016}
{Lenki{\'c}}, L., {Tzanavaris}, P., {Gallagher}, S.~C., {et~al.} 2016, \mnras,
  459, 2948, \dodoi{10.1093/mnras/stw779}

\bibitem[{{Lisenfeld} {et~al.}(2017){Lisenfeld}, {Alatalo}, {Zucker},
  {Appleton}, {Gallagher}, {Guillard}, \& {Johnson}}]{Lisenfeld2017}
{Lisenfeld}, U., {Alatalo}, K., {Zucker}, C., {et~al.} 2017, \aap, 607, A110,
  \dodoi{10.1051/0004-6361/201730898}

\bibitem[{{Lisenfeld} {et~al.}(2004){Lisenfeld}, {Braine}, {Duc}, {Brinks},
  {Charmandaris}, \& {Leon}}]{Lisenfeld2004}
{Lisenfeld}, U., {Braine}, J., {Duc}, P.~A., {et~al.} 2004, \aap, 426, 471,
  \dodoi{10.1051/0004-6361:20041330}

\bibitem[{{Lockman} {et~al.}(2012){Lockman}, {Free}, \&
  {Shields}}]{Lockman2012}
{Lockman}, F.~J., {Free}, N.~L., \& {Shields}, J.~C. 2012, \aj, 144, 52,
  \dodoi{10.1088/0004-6256/144/2/52}

\bibitem[{{Masui} {et~al.}(2013){Masui}, {Switzer}, {Banavar}, {Bandura},
  {Blake}, {Calin}, {Chang}, {Chen}, {Li}, {Liao}, {Natarajan}, {Pen},
  {Peterson}, {Shaw}, \& {Voytek}}]{Masui2013}
{Masui}, K.~W., {Switzer}, E.~R., {Banavar}, N., {et~al.} 2013, \apjl, 763,
  L20, \dodoi{10.1088/2041-8205/763/1/L20}

\bibitem[{{Michel-Dansac} {et~al.}(2010){Michel-Dansac}, {Duc}, {Bournaud},
  {Cuillandre}, {Emsellem}, {Oosterloo}, {Morganti}, {Serra}, \&
  {Ibata}}]{Michel-Dansac2010}
{Michel-Dansac}, L., {Duc}, P.-A., {Bournaud}, F., {et~al.} 2010, \apjl, 717,
  L143, \dodoi{10.1088/2041-8205/717/2/L143}

\bibitem[{{Moles} {et~al.}(1997){Moles}, {Sulentic}, \&
  {M{\'a}rquez}}]{Moles1997}
{Moles}, M., {Sulentic}, J.~W., \& {M{\'a}rquez}, I. 1997, \apjl, 485, L69,
  \dodoi{10.1086/310817}

\bibitem[{{Peterson} \& {Shostak}(1980)}]{1980ApJ...241L...1P}
{Peterson}, S.~D., \& {Shostak}, G.~S. 1980, \apjl, 241, L1,
  \dodoi{10.1086/183348}


\bibitem[{{Pontoppidan} {et~al.}(2022){Pontoppidan}, {Barrientes}, {Blome},
  {Braun}, {Brown}, {Carruthers}, {Coe}, {DePasquale}, {Espinoza}, {Marin},
  {Gordon}, {Henry}, {Hustak}, {James}, {Jenkins}, {Koekemoer}, {LaMassa},
  {Law}, {Lockwood}, {Moro-Martin}, {Mullally}, {Pagan}, {Player}, {Proffitt},
  {Pulliam}, {Ramsay}, {Ravindranath}, {Reid}, {Robberto}, {Sabbi}, {Ubeda},
  {Balogh}, {Flanagan}, {Gardner}, {Hasan}, {Meinke}, \&
  {Nota}}]{Pontoppidan2022}
{Pontoppidan}, K.~M., {Barrientes}, J., {Blome}, C., {et~al.} 2022, \apjl, 936,
  L14, \dodoi{10.3847/2041-8213/ac8a4e}

\bibitem[{{Qian} {et~al.}(2020){Qian}, {Yao}, {Sun}, {Xu}, {Pan}, \&
  {Jiang}}]{2020Innov...100053Q}
{Qian}, L., {Yao}, R., {Sun}, J., {et~al.} 2020, The Innovation, 1, 100053,
  \dodoi{10.1016/j.xinn.2020.100053}

\bibitem[{{Renaud} {et~al.}(2010){Renaud}, {Appleton}, \& {Xu}}]{Renaud2010}
{Renaud}, F., {Appleton}, P.~N., \& {Xu}, C.~K. 2010, \apj, 724, 80,
  \dodoi{10.1088/0004-637X/724/1/80}

\bibitem[{{Renaud} {et~al.}(2018){Renaud}, {Athanassoula}, {Amram}, {Bosma},
  {Bournaud}, {Duc}, {Epinat}, {Fensch}, {Kraljic}, {Perret}, \&
  {Struck}}]{Renaud2018}
{Renaud}, F., {Athanassoula}, E., {Amram}, P., {et~al.} 2018, \mnras, 473, 585,
  \dodoi{10.1093/mnras/stx2360}

\bibitem[{{Sanders} \& {Mirabel}(1996)}]{Sanders1996}
{Sanders}, D.~B., \& {Mirabel}, I.~F. 1996, \araa, 34, 749,
  \dodoi{10.1146/annurev.astro.34.1.749}

\bibitem[{{Shostak} {et~al.}(1980){Shostak}, {Gilra}, {Noordam},
  {Nieuwenhuijzen}, {Vermue}, \& {de Graauw}}]{Shostak1980}
{Shostak}, G.~S., {Gilra}, D.~P., {Noordam}, J.~E., {et~al.} 1980, \aap, 81,
  223

\bibitem[{{Shostak} {et~al.}(1984){Shostak}, {Sullivan}, \&
  {Allen}}]{Shostak1984}
{Shostak}, G.~S., {Sullivan}, W.~T., I., \& {Allen}, R.~J. 1984, \aap, 139, 15

\bibitem[{{Stephan}(1877)}]{Stephan1877}
{Stephan}, M. 1877, \mnras, 37, 334, \dodoi{10.1093/mnras/37.6.334}

\bibitem[{{Sulentic} {et~al.}(2001){Sulentic}, {Rosado}, {Dultzin-Hacyan},
  {Verdes-Montenegro}, {Trinchieri}, {Xu}, \& {Pietsch}}]{Sulentic2001}
{Sulentic}, J.~W., {Rosado}, M., {Dultzin-Hacyan}, D., {et~al.} 2001, \aj, 122,
  2993, \dodoi{10.1086/324455}

\bibitem[{{Tammann}(1970)}]{Tammann1970}
{Tammann}, G.~A. 1970, \aplett, 7, 111

\bibitem[{{Tepper-Garc{\'\i}a} {et~al.}(2022){Tepper-Garc{\'\i}a},
  {Bland-Hawthorn}, \& {Freeman}}]{Tepper-Garcia2022}
{Tepper-Garc{\'\i}a}, T., {Bland-Hawthorn}, J., \& {Freeman}, K. 2022, \mnras,
  515, 5951, \dodoi{10.1093/mnras/stac1926}

\bibitem[{{Trinchieri} {et~al.}(2005){Trinchieri}, {Sulentic}, {Pietsch}, \&
  {Breitschwerdt}}]{Trinchieri2005}
{Trinchieri}, G., {Sulentic}, J., {Pietsch}, W., \& {Breitschwerdt}, D. 2005,
  \aap, 444, 697, \dodoi{10.1051/0004-6361:20052910}

\bibitem[{{Umehata} {et~al.}(2019){Umehata}, {Fumagalli}, {Smail}, {Matsuda},
  {Swinbank}, {Cantalupo}, {Sykes}, {Ivison}, {Steidel}, {Shapley}, {Vernet},
  {Yamada}, {Tamura}, {Kubo}, {Nakanishi}, {Kajisawa}, {Hatsukade}, \&
  {Kohno}}]{Umehata2019}
{Umehata}, H., {Fumagalli}, M., {Smail}, I., {et~al.} 2019, Science, 366, 97,
  \dodoi{10.1126/science.aaw5949}

\bibitem[{{van der Hulst} \& {Rots}(1981)}]{vanderhulst1981}
{van der Hulst}, J.~M., \& {Rots}, A.~H. 1981, \aj, 86, 1775,
  \dodoi{10.1086/113060}

\bibitem[{{Verdes-Montenegro} {et~al.}(2001){Verdes-Montenegro}, {Yun},
  {Williams}, {Huchtmeier}, {Del Olmo}, \& {Perea}}]{VerdesMontenegro2001}
{Verdes-Montenegro}, L., {Yun}, M.~S., {Williams}, B.~A., {et~al.} 2001, \aap,
  377, 812, \dodoi{10.1051/0004-6361:20011127}

\bibitem[{{Walker} {et~al.}(2019){Walker}, {Simionescu}, {Nagai}, {Okabe},
  {Eckert}, {Mroczkowski}, {Akamatsu}, {Ettori}, \& {Ghirardini}}]{Walker2019}
{Walker}, S., {Simionescu}, A., {Nagai}, D., {et~al.} 2019, \ssr, 215, 7,
  \dodoi{10.1007/s11214-018-0572-8}

\bibitem[{{Williams} {et~al.}(2002){Williams}, {Yun}, \&
  {Verdes-Montenegro}}]{Williams2002}
{Williams}, B.~A., {Yun}, M.~S., \& {Verdes-Montenegro}, L. 2002, \aj, 123,
  2417, \dodoi{10.1086/339839}

\bibitem[{{Wolfe} {et~al.}(2016){Wolfe}, {Lockman}, \& {Pisano}}]{Wolfe2016}
{Wolfe}, S.~A., {Lockman}, F.~J., \& {Pisano}, D.~J. 2016, \apj, 816, 81,
  \dodoi{10.3847/0004-637X/816/2/81}

\bibitem[{{Xu} {et~al.}(1999){Xu}, {Sulentic}, \& {Tuffs}}]{Xu1999}
{Xu}, C., {Sulentic}, J.~W., \& {Tuffs}, R. 1999, \apj, 512, 178,
  \dodoi{10.1086/306771}

\bibitem[{{Xu} {et~al.}(2021){Xu}, {Lisenfeld}, {Gao}, \& {Renaud}}]{XU2021}
{Xu}, C.~K., {Lisenfeld}, U., {Gao}, Y., \& {Renaud}, F. 2021, \apj, 918, 55,
  \dodoi{10.3847/1538-4357/ac0f77}

\bibitem[{{Xu} {et~al.}(2003){Xu}, {Lu}, {Condon}, {Dopita}, \&
  {Tuffs}}]{Xu2003}
{Xu}, C.~K., {Lu}, N., {Condon}, J.~J., {Dopita}, M., \& {Tuffs}, R.~J. 2003,
  \apj, 595, 665, \dodoi{10.1086/377445}

\bibitem[{{Xu} {et~al.}(2005){Xu}, {Iglesias-P{\'a}ramo}, {Burgarella}, {Rich},
  {Neff}, {Lauger}, {Barlow}, {Bianchi}, {Byun}, {Forster}, {Friedman},
  {Heckman}, {Jelinsky}, {Lee}, {Madore}, {Malina}, {Martin}, {Milliard},
  {Morrissey}, {Schiminovich}, {Siegmund}, {Small}, {Szalay}, {Welsh}, \&
  {Wyder}}]{Xu2005}
{Xu}, C.~K., {Iglesias-P{\'a}ramo}, J., {Burgarella}, D., {et~al.} 2005, \apjl,
  619, L95, \dodoi{10.1086/425130}

\bibitem[{{Xu} {et~al.}(2022){Xu}, {Cheng}, {Appleton}, {Duc}, {Gao}, {Tang},
  {Yun}, {Dai}, {Huang}, {Lisenfeld}, \& {Renaud}}]{Xu2022}
{Xu}, C.~K., {Cheng}, C., {Appleton}, P.~N., {et~al.} 2022, \nat, 610, 461,
  \dodoi{10.1038/s41586-022-05206-x}

\bibitem[{{Yttergren} {et~al.}(2021){Yttergren}, {Misquitta},
  {S{\'a}nchez-Monge}, {Valencia-S}, {Eckart}, {Zensus}, \&
  {Peitl-Thiesen}}]{Yttergren2021}
{Yttergren}, M., {Misquitta}, P., {S{\'a}nchez-Monge}, {\'A}., {et~al.} 2021,
  \aap, 656, A83, \dodoi{10.1051/0004-6361/202040188}

\bibitem[{{Yun} {et~al.}(1997){Yun}, {Verdes-Montenegro}, {del Olmo}, \&
  {Perea}}]{Yun1997}
{Yun}, M.~S., {Verdes-Montenegro}, L., {del Olmo}, A., \& {Perea}, J. 1997,
  \apjl, 475, L21, \dodoi{10.1086/310458}

\bibitem[{{Zwicky} \& {Kowal}(1968)}]{ZwickyKowal1968}
{Zwicky}, F., \& {Kowal}, C.~T. 1968, {``Catalogue of Galaxies and of Clusters
  of Galaxies'', Volume VI} (Pasadena: California Institute of Technology)

\end{thebibliography}
\end{document}